\documentclass[journal]{IEEEtran}
\ifCLASSINFOpdf
  \usepackage[pdftex]{graphicx}
  \graphicspath{../eps/}
  \DeclareGraphicsExtensions{.pdf,.jpeg,.png}
\else
  \usepackage[dvips]{graphicx}
  \graphicspath{{../eps/}}
  \DeclareGraphicsExtensions{.eps}
\fi
\usepackage{siunitx}
\usepackage[colorlinks=false]{hyperref}
\usepackage{color}
\usepackage{cite}
\usepackage{amsmath}
\usepackage{amssymb}
\usepackage{mathptmx}
\usepackage{epsfig}
\usepackage{times}
\usepackage{threeparttable}
\usepackage{stfloats}
\usepackage{amsmath,amssymb,amsfonts}
\usepackage{algorithmic}
\usepackage{textcomp}
\usepackage{array}
\usepackage{multirow}
\usepackage{array}

  \usepackage[caption=false,font=footnotesize]{subfig}
\begin{document}
%
\title{ASCH-PUF: A ``Zero" Bit Error Rate \\CMOS Physically Unclonable Function \\with Dual-Mode Low-Cost Stabilization }
%
%
%

\author{
        Yan He,~\IEEEmembership{Graduate Student Member,~IEEE,}
        Dai Li,
        Zhanghao Yu,~\IEEEmembership{Graduate Student Member,~IEEE,}
        and~Kaiyuan~Yang,~\IEEEmembership{Member,~IEEE}
        \thanks{Manuscript received on}
\thanks{Y. He, D. Li, Z. Yu, and K. Yang are with the Department of Electrical and Computer Engineering, Rice University, Houston TX, 77005, USA.}
\thanks{(Corresponding Author: Kaiyuan Yang, kyang@rice.edu)}
\thanks{1063-8210 © 2023 IEEE. Personal use is permitted, but republication/redistribution
requires IEEE permission.}
}

%
%

\markboth{IEEE Journal of Solid-State Circuits}%
{Shell \MakeLowercase{\textit{et al.}}: Bare Demo of IEEEtran.cls for IEEE Journals}
%



\maketitle

\begin{abstract}
Physically unclonable functions (PUFs) are increasingly adopted for low-cost and secure secret key and chip ID generations for embedded and IoT devices. Achieving 100\% reproducible keys across wide temperature and voltage variations over the lifetime of a device is critical and conventionally requires large masking or Error Correction Code (ECC) overhead to guarantee. This paper presents an Automatic Self Checking and Healing (ASCH) stabilization technique for a state-of-the-art PUF cell design based on sub-threshold inverter chains. 
The ASCH system successfully removes all unstable PUF cells without the need for expensive temperature sweeps during unstable bit detection.
By accurately finding all unstable bits without expensive temperature sweeps to find all unstable bits, ASCH achieves ultra-low bit error rate (BER), thus significantly reducing the costs of using ECC and enrollment. 
Our ASCH can operate in two modes, a static mode (S-ASCH) with a conventional pre-enrolled unstable bit mask and a dynamic mode (D-ASCH) that further eliminates the need for non-volatile memories (NVMs) for storing masks. 
The proposed ASCH-PUF is fabricated and evaluated in 65nm CMOS. 
The ASCH system achieves ``0" Bit Error Rate (BER, \textless 1.77E-9) across temperature variations of -20°C to 125°C, and voltage variations of 0.7V to 1.4V, by masking 31\% and 35\% of all fabricated PUF bits in S-ASCH and D-ASCH mode respectively. 
The prototype achieves a measured throughput of 11.4 Gbps with 0.057 fJ/b core energy efficiency at 1.2V, 25°C.

\end{abstract}


%
\IEEEpeerreviewmaketitle

\section{Introduction}
Security is a major concern in the Internet of Things (IoT) devices due to cost, energy, and computational resource constraints, as well as vulnerabilities to various physical attacks~\cite{yang2017security}. The secret keys used for authentication and encryption are the root of trust, demanding superb security without compromising other hardware metrics. Traditionally, None-Volatile Memory (NVM) or One-Time Programmable (OTP) memories are used to store the permanent keys. This method is susceptible to probing and optical attacks and is costly both due to its large chip area and the additional fabrication steps. Physically Unclonable Function (PUF) is the most promising alternative for secure key storage. A PUF generates random and unique keys for each device by extracting the unique intrinsic process variations. The secret key is generated only when the PUF is powered on, eliminating its vulnerability to physical attacks. Any attempt to steal the key will likely disturb the subtle variations in PUF circuits, making PUF tamper-evident. Lastly, PUFs require significantly less power and area overheads than NVM-based solutions.


Fabricated PUF chips initially go through an enrollment process at nominal conditions, where every bit is collected as the golden values. Then the chips will be deployed and interrogated in the field for key generations. The stability, or reproducibility, of PUF values under environmental variations (e.g., noise, voltage, temperature, or aging) is the critical challenge for silicon PUFs that rely on small process variations of CMOS transistors. 
The metric used to evaluate the stability of a PUF is Bit Error Rate (BER). 
For evaluating the BER, each PUF bit is read out and compared against its golden value, leading to a metric defined as,
\begin{equation}
\label{eq:BER}
    B E R=\frac{\# \text { Bit Error }}{\# \text { Bit } \times \# \text { Evaluation }}
\end{equation}
A PUF must meet the Key Error Rate (KER) of its cryptographic application. The KER is the possibility of having at least one-bit error in an N-bit key, i.e., $KER={1-(1-BER)^N}$.
For a 128-bit key with a typical KER requirement of 1E-6 \cite{taneja2021puf}, the required PUF BER is around 1E-8, which is much lower than that of a state-of-art PUF’s native BER \cite{liu5VHybridSRAM2021,li2019self,liu2020373,choi2020physically,taneja2018fully,karpinskyy20168,lee2018445f,satpathy20174,li2016ultra,yang20178,park2022ber}.

\begin{figure}[t]
      \centering
      \includegraphics[scale=0.76]{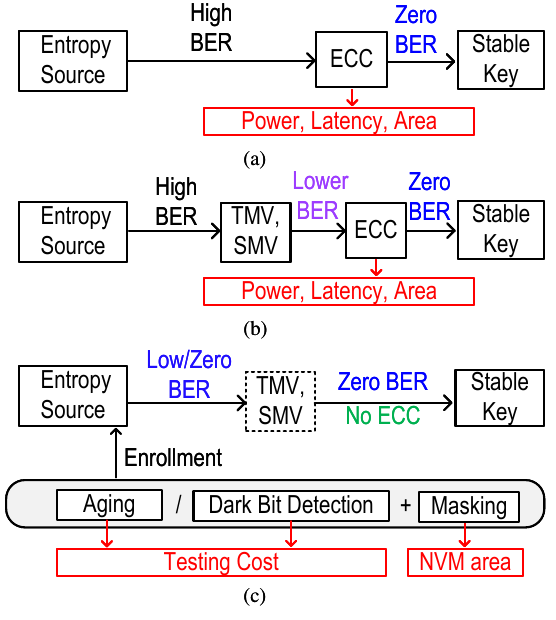}
      \vskip -2ex
      \caption{Existing PUF stabilization methods: (a) ECC, (b) TMV/SMV with ECC, and (c) stabilization during enrollment. }
      \label{intro}
  \end{figure}

Error Correction Code (ECC), such as BCH code \cite{mathew201416,karpinskyy20168}, is studied to help any PUF design achieve the desired BER (see Fig.~\ref{intro}a). But the area, power, and latency overheads of ECC scale super-linearly with BER~\cite{li2019self,basak2013reconfigurable}. Thus, low-overhead stabilization of PUFs before ECC is crucial.

Run-time stabilization, like Temporal/Spatial Majority Voting (TMV/SMV) \cite{mathew201416}, can effectively remove the bit errors induced by random noise (see Fig.~\ref{intro}b). But they can hardly mitigate the voltage/temperature (V/T) effects that are the dominant source of PUF's instability. 
Intentional burn-in of PUF cells is another effective BER reduction method by hardening the native PUF circuit variations through accelerated aging. 
\cite{satpathy20174} improves BER by 50\% with virtually no design overhead while \cite{liu5VHybridSRAM2021}
achieves ``0" BER after burn-in using hot carrier injection (HCI). The downside of this method is that every chip is subject to an accelerated aging process, during which a higher-than-nominal supply voltage and/or temperature is applied to stress the PUF transistor, incurring additional testing costs. 

\begin{figure}[t]
      \centering
      \includegraphics[scale=0.4]{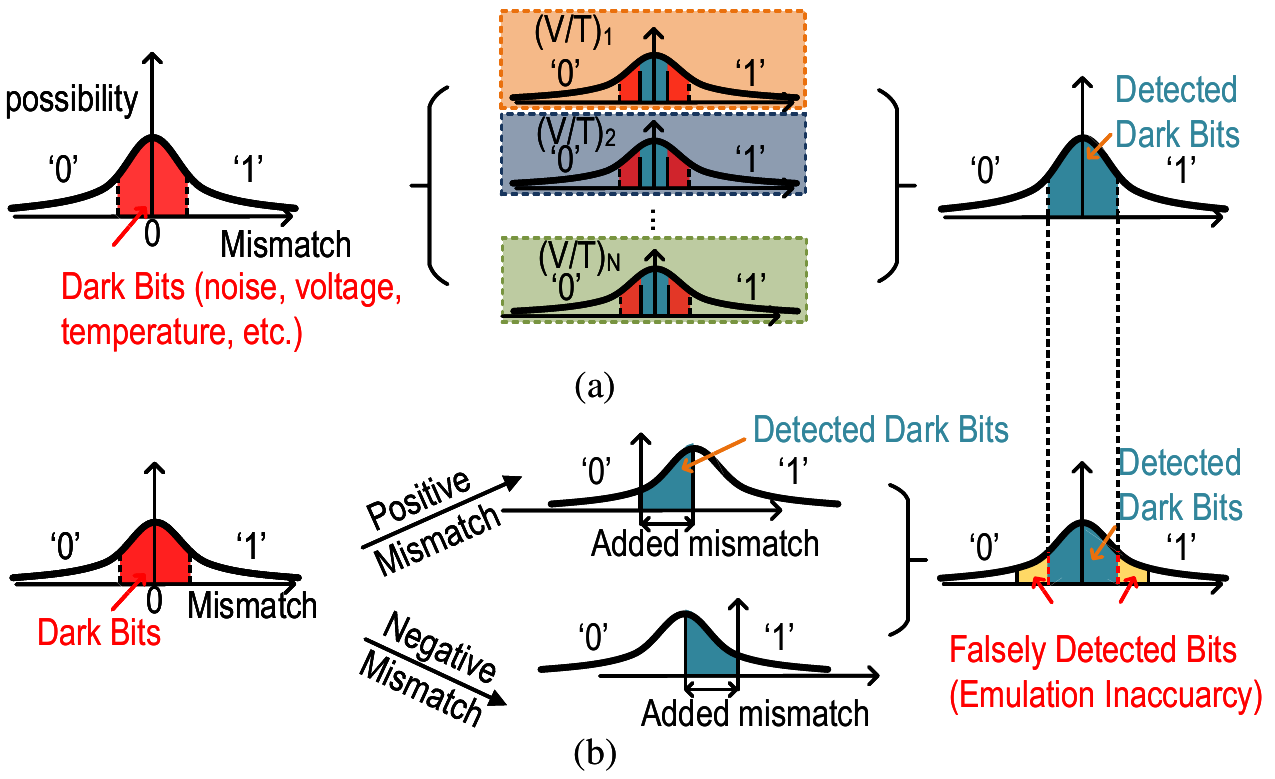}
      \vskip -2ex
      \caption{(a) Conventional dark bit detection scheme, and (b) dark bit detection through V/T variation emulation.}
      \label{DBD}
  \end{figure}

On the other hand, the masking technique reduces BER by finding and replacing the potentially unstable ``dark" bits that may flip under V/T variations (see Fig.~\ref{intro}c). 
Conventionally, as shown in Fig.~\ref{DBD}a, all PUF bits will be evaluated many times under different V/T corners. The mapping of dark bits is stored in NVM for masking in normal operations. Because PUFs are tested under every possible condition, this dark bit detection method is highly accurate but requires high testing efforts.

In order to leverage the effectiveness of hard masking without the testing costs, several designs have been proposed to emulate the V/T variation by adding a positive and negative mismatch to the PUF circuit~\cite{yang20178,li2019self,liu2020373,shifman2020sram} (see Fig.~\ref{DBD}b). This approach drastically reduces testing costs, but the emulation inaccuracy leads to falsely flagged stable bits and an unnecessarily high masking ratio (defined as the percentage of masked PUF cells).
Using this method, some designs even achieve ``0" BER~\cite{liu2020373,shifman2020sram}. Since it is unreasonable to claim a PUF will never have an error, the BER calculation is commonly modified to pessimistically assume that the next evaluation of PUF will be an error bit. Thus the BER of PUFs with no observed error in testing is defined as,
\begin{equation} \label{eq:pessimisticBER}
B E R_{pes}=\frac{1}{\# \text {Bit} \times(1-\text {Masking Ratio}) \times \# \text {Evaluation}}
\end{equation}
where (1-Masking Ratio) means only the unmasked bits are included in the BER calculation.

In \cite{shifman2020sram}, modulating the capacitive load of individual inverters inside SRAM PUF adds a mismatch in either direction. By masking 59\% of the total PUF bits, it achieved ``0" BER of $<$1E-9 under a restricted temperature range (-10\textdegree C to 85\textdegree C). In \cite{liu2020373}, the SRAM mismatch is shifted in two directions by adding ground resistance in either inverter, and “0” BER of $<$5.99E-7 is achieved with a 61\% Masking Ratio across a wide temperature range. But an additional dark bit detection process with voltage overdrive to 1.6V is needed to correct the voltage variation at the worst corner, which may not be feasible in newer process nodes with low voltage compliance. 
Even though these designs reduce or eliminate the ECC requirements, they failed to address the overheads associated with aggressive hard masking with a high masking ratio. 
In prior arts, due to the low detection accuracy, more than half of the PUF array is masked to achieve ``0" BER. 
Such high masking ratios result in a large NVM overhead and require high redundancy for error recovery. These overheads diminish the claim of PUF being lower costs than NVM solutions and make PUF less attractive in resource-constraint applications.

\cite{satpathy201413fj} proposed a soft masking approach to alleviating the conventional masking overheads. A flip detection circuit is added to the PUF readout to determine if a cell is “valid” in run-time. Instead of discarding and replacing the unstable bits, they are directly set to one or zero. This run-time soft masking circumvents NVM storage and pre-testing, but it has its own limitations, including 1) forcing unstable bits to a set value compromises the randomness of PUF, 2) unpredictability of keys because the generated run-time mask may vary over time and environments, and 3) inability to capture all V/T variations because the masking is only based on run-time information without comparison to the golden enrollment values. 
Despite these restrictions, the run-time valid detection approach proposed in the paper is very interesting and is adopted in different forms (\cite{taneja2021puf,taneja2018fully,lee2021samsung}) because it enables the output of real-time stability information without reading out the PUF value, which is attractive both from a security perspective and for the purpose of automation. 

This paper proposes a ``0" BER PUF array exploring an Automatic Self Checking and Healing (ASCH) stabilization scheme with dual operation modes. The choice of operation modes is application specific, depending on the desired trade-offs between device area and server capacity. This work makes the following contributions.

\begin{figure}[t]
      \centering
      \includegraphics[width=0.85\columnwidth]{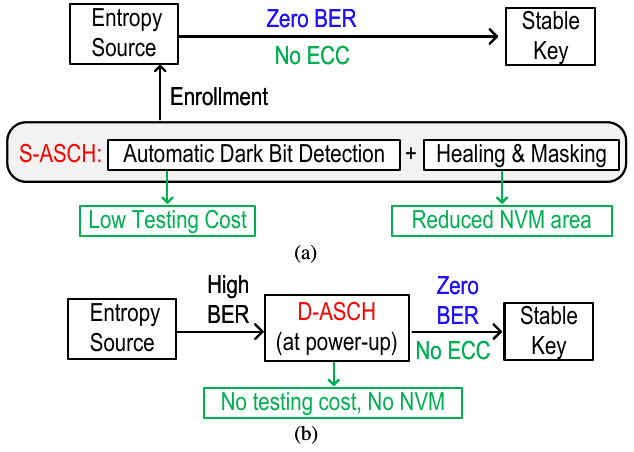}
      \vskip -2ex
      \caption{Proposed PUF stabilization methods (a) S-ASCH and (b) D-ASCH.}
      \label{PUFintro_ASCH}
  \end{figure}

\begin{itemize}
    \item A novel high-precision dark bit detection and healing approach (ASCH) greatly minimizes the masking ratio to achieve ``0" BER after fast, automated, and low-cost testing. 
    
    \item Static mode of ASCH (S-ASCH) (Fig. \ref{PUFintro_ASCH}a) enables stabilization during enrollment, which achieves ``0" BER of \textless1.77E-09 with merely 31\% masking ratio, requiring only \textless4 ms testing time for a 4096-bit array (NVM write-in time not included).
    
	\item Dynamic ASCH (D-ASCH) (Fig. \ref{PUFintro_ASCH}b) achieves comparable stabilization effects while replacing the NVM storage with in-field dark bit detection. The soft masking principle is exploited in D-ASCH with modified stabilization and communication protocols to circumvent its limitations~\cite{satpathy201413fj}.
    
    \item A 65nm CMOS ASCH-PUF prototype demonstrates high resistance to voltage variations, ultra-compact footprint, ultra-low-power consumption, and high throughput. The entire ASCH mechanism is highly digital and process node scalable.
\end{itemize}

The remainder of this article is organized as follows. Section II describes the proposed ASCH-PUF with high-precision self-checking and healing. Section III explains the implementation and trade-off of the two stabilization modes. In Section IV, measurement results are presented and discussed. A conclusion is drawn in Section V.

\section{In-Situ Dark Bit Checking and Healing}
The proposed dual-mode stabilization system relies on a fast, accurate, low-cost self-checking and healing scheme integrated within the PUF array. Self-checking finds the dark bits with a configurable threshold, and the healing step reduces the number of dark bits through in-cell reconfiguration. 

\subsection{Dark Bit Detection}
\label{subsec:Checking}
\begin{figure}[t]
      \centering
      \includegraphics[width=0.85\columnwidth]{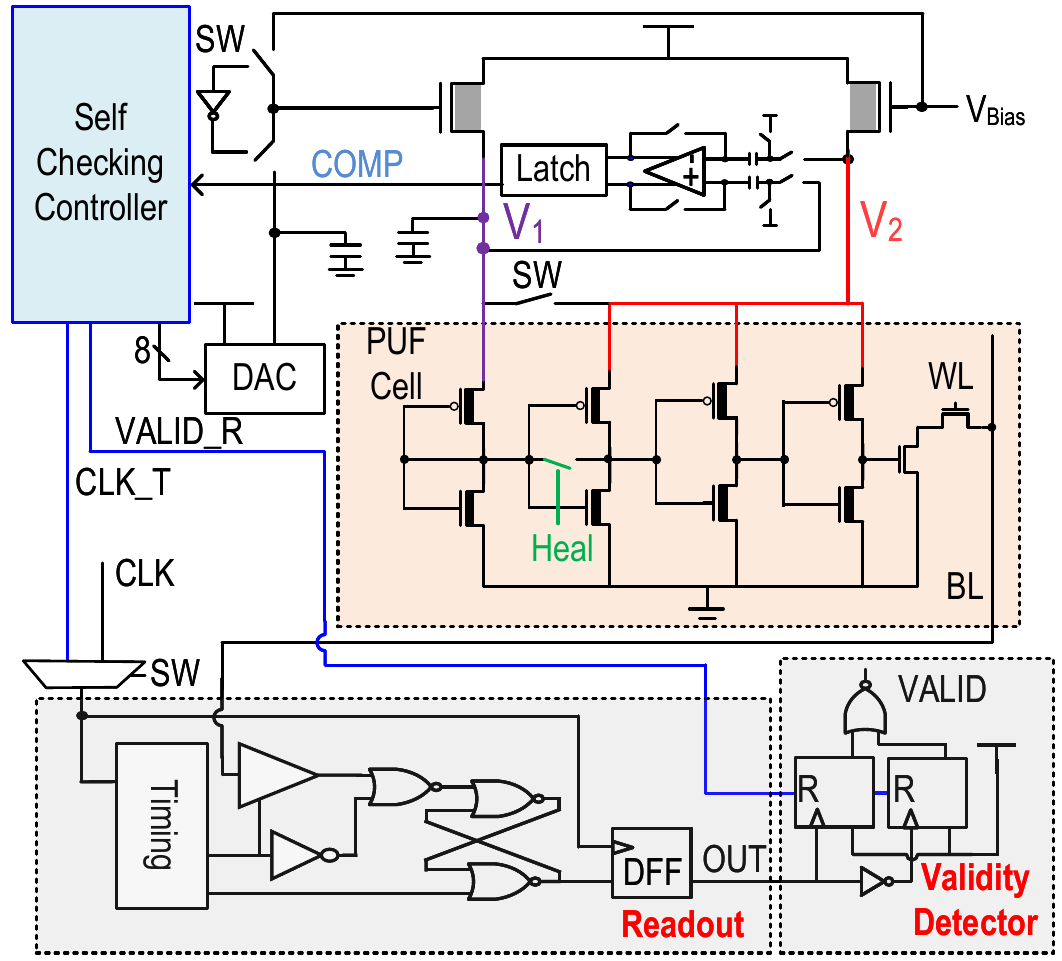}
      \vskip -2ex
      \caption{Schematic of self-checking and healing on a single PUF cell, with the implementation of auto-zero comparator, readout circuit, and validity detector.}
      \label{SCsystem}
  \end{figure}

The high-precision dark bit self-checking function is enabled by the PUF system shown in Fig.~\ref{SCsystem}.
The PUF cell is an inverter-based PUF \cite{yang20178} that demonstrated ultra-low power consumption, state-of-the-art native stability, and a compact footprint. It utilizes a native transistor header as a low-cost regulation solution, which enables the sub-threshold operation of a PUF cell and improves its resistance to voltage variations. 

\begin{figure}[t]
      \centering
      \includegraphics[scale=0.43]{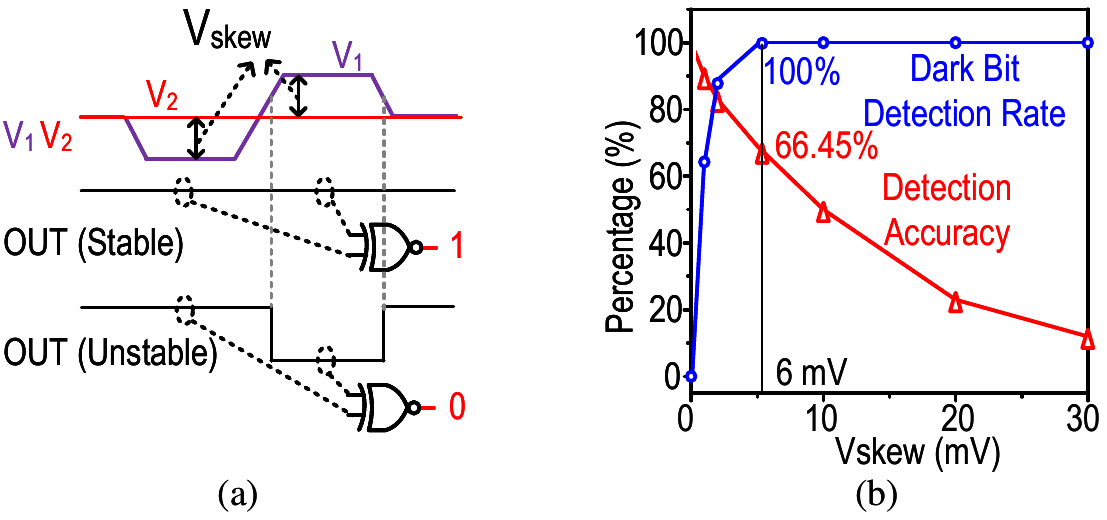}
      \vskip -2ex
      \caption{(a) Principle of voltage skew and dark bit detection, and (b) simulation result of dark bit detection showing the relationship between Detection Accuracy and Detection Rate.}
      \label{Dsim}
  \end{figure}

In order to emulate the change of $V_{th}$ due to different V/T variations without changing the actual condition, a source of mismatch that tilts the PUF in both directions is required. In the case of inverter-based PUF, the change of $V_{th}$ manifests as the change in the difference of switching voltage between the first-stage and second-stage inverter. In the proposed design, this change of switching voltage is emulated by controlling the first stage inverter supply voltage, $V_{1}$, using an 8-bit resistive DAC. Change in both directions is achieved by first decreasing and then increasing the DAC value, which leads to a negative and positive voltage skew, as shown in Fig.~\ref{Dsim}. The absolute value of the skewing voltage is kept the same in both directions to represent the same amount of mismatch. If the PUF output from two opposite skews is flipped, this PUF cell is deemed as a dark bit.
Spice simulation shows that the design has 179 unstable bits out of 100000, representing 1.79\% instability across -40 to 125 degrees. At 6 mV $V_{skew}$, all unstable bits are detected, achieving a 66.45\% detection accuracy. The detection accuracy is defined as
\begin{equation}
\text { Detection Accuracy }=\frac{\text { \# of Unstable Bits }}{\# \text { of Dark Bits }}
\end{equation}
Higher detection accuracy means a reduced number of falsely rejected bits, which leads to a lower masking ratio. The simulation result shows a good correlation between the mismatch and actual V/T variation.

An SRAM-like peripheral is integrated for high-speed parallel readout. The circuit samples and output PUF value at every rising clock edge. The Validity Detector is used for the automatic detection of unstable cells by checking if a PUF is stable during its evaluation window. It functions by de-asserting the Reset during evaluation and outputs a "1" from either D-FF if there is a PUF transition. The NORed output is ``0" if the evaluated bit is unstable, and vice versa.

\begin{figure}[t]
      \centering
      \includegraphics[scale=0.5]{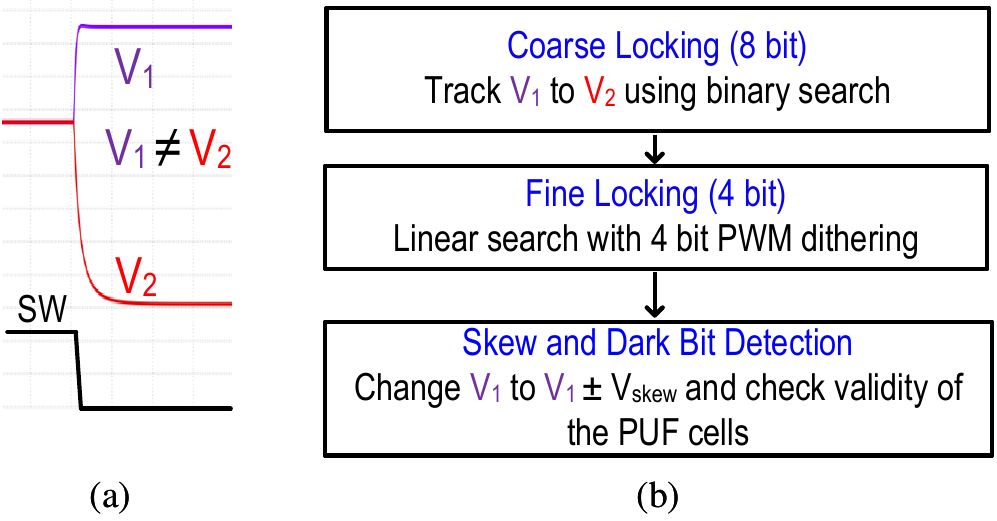}
      \vskip -2ex
      \caption{(a) Load imbalance issue when SW is turned to 0. (b) Self-checking with coarse-fine Locking.}
      \label{Locking}
  \end{figure}

The self-checking process is challenging because of the load imbalance issue. As shown in Fig.~\ref{SCsystem}, during normal operation, both native regulators are biased by an external source $V_{bias}$, and $V_{1}$ is shorted with $V_{2}$. At the start of the checking operation, the native regulator for $V_{1}$ is biased by DAC, and $V_{1}$ is isolated from $V_{2}$. As shown in Fig.~\ref{Locking} (a), This results in an imbalance between $V_{1}$ and $V_{2}$ due to the different loading conditions. To detect the unstable bits caused by the same amount of intentional mismatch in both directions, $V_{1}$ value needs to be tuned close to $V_{2}$ before adding voltage skew.

\begin{figure}[t]
      \centering
      \includegraphics[scale=0.5]{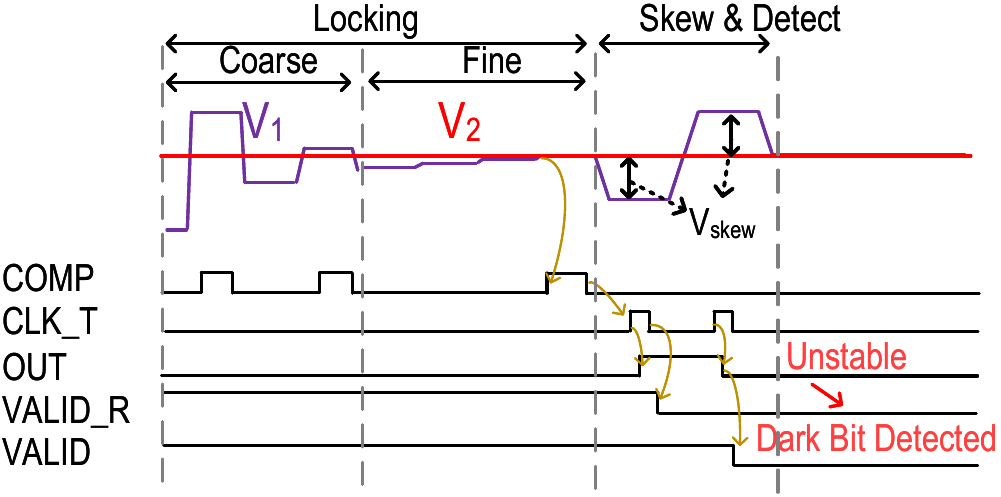}
      
      \caption{Illustrated self-checking waveform, demonstrating the detection of an unstable PUF cell.}
      \label{LockingWaveform_Ill}
  \end{figure}

As shown in Fig.~\ref{Locking} (b), a coarse-fine locking process is added before skew and detection to enable self-checking. During coarse locking, the 8-bit DAC searches for the closest $V_{1}$ value to $V_{2}$. The DAC value is dithered using 4-bit pulse width modulation. Capacitor $C_{1}$ and $C_{2}$ are added to stabilize the dithered voltage. This fine step locking increases the resolution to 12 bits. During locking, a sufficient delay is provided to stabilize voltage after each step change. An auto-zeroing comparator is activated 5 times at the end of each step, and a majority of the result is used for accurate comparison between $V_{1}$ and $V_{2}$. The digitized $V_{1}$ value is saved in the controller after locking, and the skew operation is performed with the same 12-bit precision. The amount of skew is provided externally. two consecutive PUF evaluation sessions under programmable skews are performed. All cells that ever flip once during the two sessions are marked as unstable by the validity detector. The number of PUF evaluations during checking affects the detection of unstable bits caused by noise, which is a smaller but not negligible source of mismatch compared with added skew. 64 evaluations for each session are decided to balance the detection accuracy and checking speed. The illustrated waveform of the self-checking process is shown in Fig.~\ref{LockingWaveform_Ill}.

\subsection{Dark Bit Healing}
\label{subsec:Healing}
After the self-checking operation, instead of directly masking all the potentially unstable bits, The proposed system further leverages the cell reconfiguration design in \cite{li2019self} to heal a large portion of unstable bits locally.

    \begin{figure}[t]
      \centering
      \includegraphics[scale=0.5]{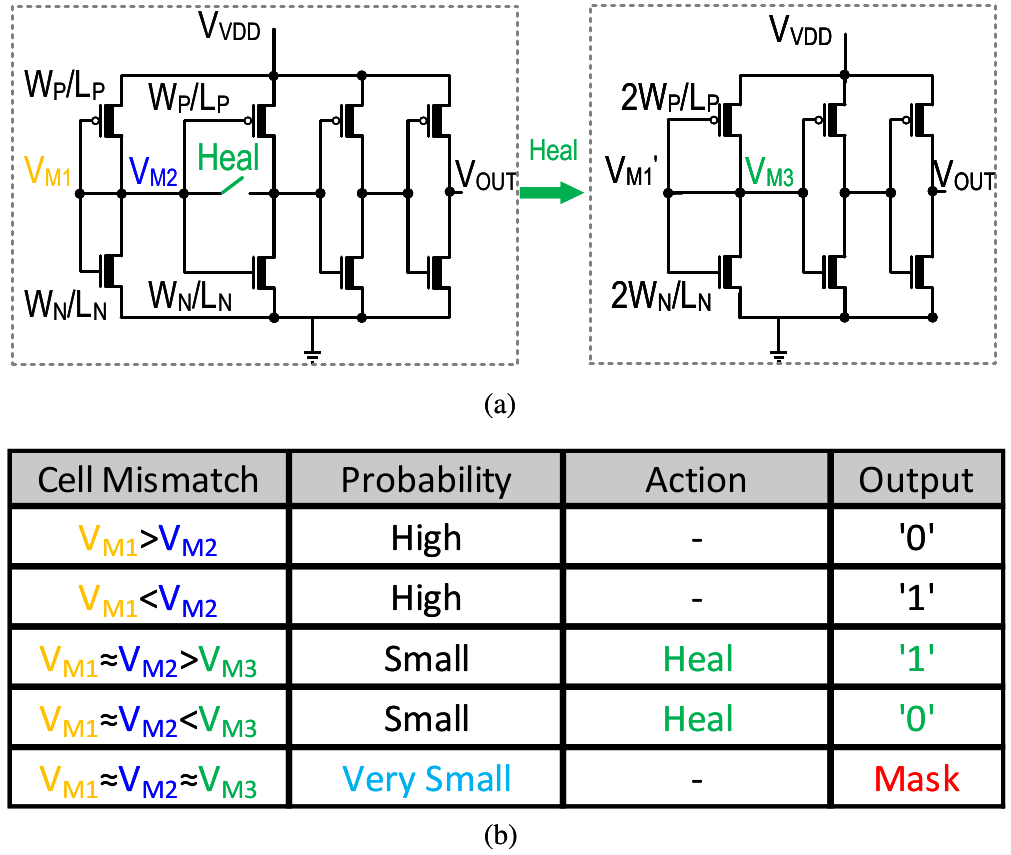}
      \vskip -2ex
      \caption{(a) Process of healing in transistor level schematic. (b) Healing condition, probability, and final output.}
      \label{Healing}
  \end{figure}
  
As shown in Fig.~\ref{Healing}, shorting stages 1 and 2 converts the original 4-stage cell into an almost independent 3-stage one. Because of the naturally small BER of the PUF cell, the probability that the 4-stage cell is unstable is already small, and the probability that both cells are unstable is even smaller. Since we only mask the cell that is unstable in both configurations, the healing operation greatly improves the masking ratio. 

The reconfigured PUF cell \cite{li2019self} exhibits a bias, which is caused by the asymmetry of the drain/source area between the third-stage inverter and the combined first and second stage. With a carefully designed layout considering the area symmetries, close-to-ideal uniqueness and identifiability are achieved for both original and heal cells, as evidenced by their respective inter- and intra- Hamming distance measurements shown in Section IV.

    \begin{figure}[t]
      \centering
      \includegraphics[width=\columnwidth]{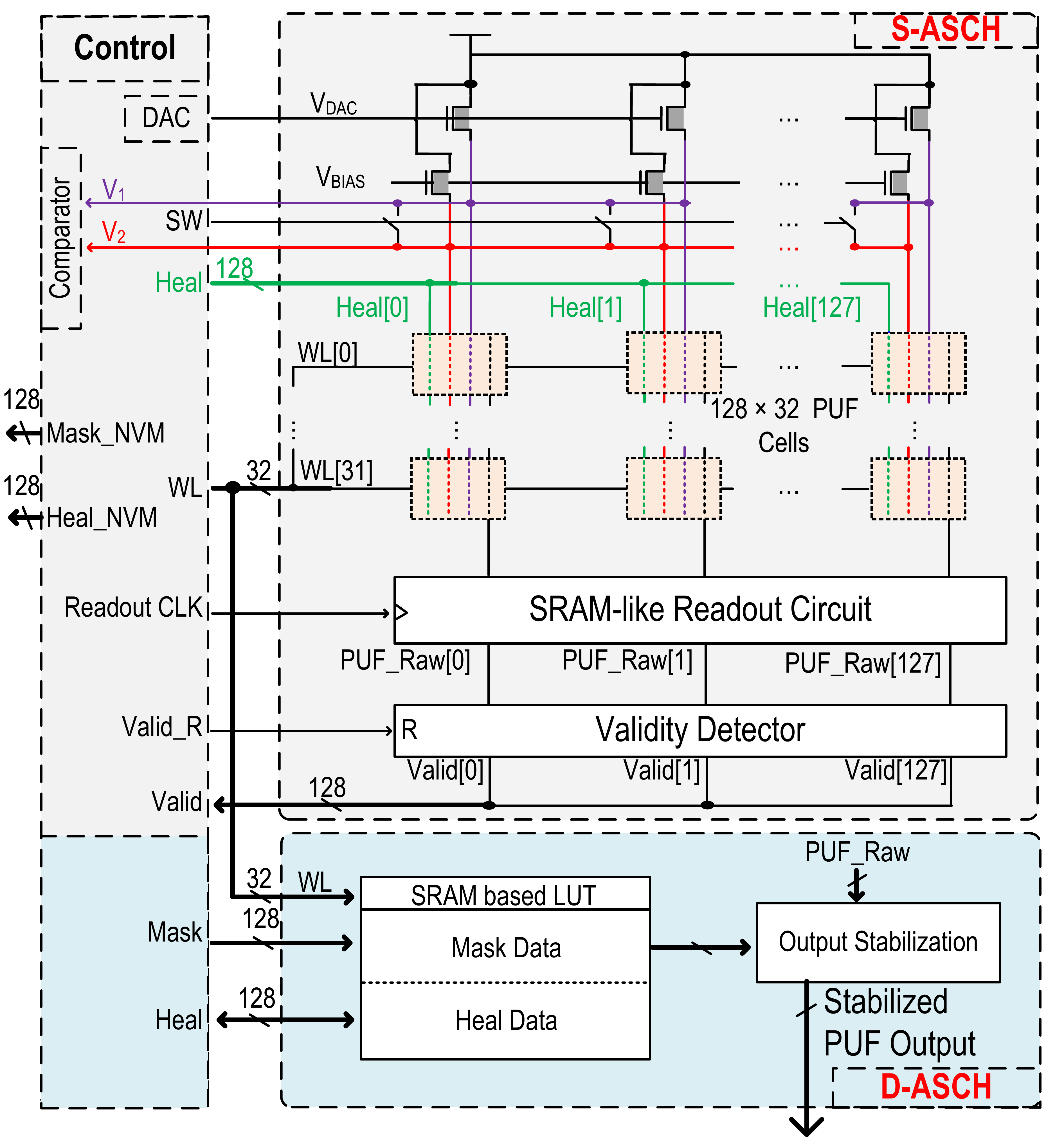}
      \vskip -2ex
      \caption{S-ASCH system diagram and the additional blocks for enabling D-ASCH operation.}
      \label{System}
  \end{figure}

\section{Dual Stabilization Modes in ASCH-PUF}
The proposed self-checking and healing scheme can be operated in two modes: Static Automatic Self Checking and Healing (S-ASCH) enables stabilization during enrollment~\cite{he202136}, and its NVM-less, dynamic alternative (D-ASCH) checks and heals the cell in-field. Both modes are capable of reducing the BER to 0 in order to eliminate the need for ECC, and the mode selection depends on the preference of the trade-off between on-chip and server storage.
The dual-mode stabilization system diagram is shown in Fig.~\ref{System}. A 128x32 PUF Array is implemented. S-ASCH takes full advantage of the proposed self-checking and healing system shown in Fig~\ref{SCsystem}. NVM for mask storage is not displayed in this figure. D-ASCH replaces the NVM with SRAM, and an additional output stabilization block is added with a small design overhead.
The detailed implementations and trade-offs of these two modes of operation are explained in this section.

\subsection{Static Automatic Self Checking and Healing (S-ASCH)}
S-ASCH automatically detects all dark bits within a V/T range and reduces BER to 0 by generating the healing and masking information. It is an extremely fast process that is run only once during enrollment, so almost no overhead is added aside from NVM implementation. 

    \begin{figure}[t]
      \centering
      \includegraphics[width=\columnwidth]{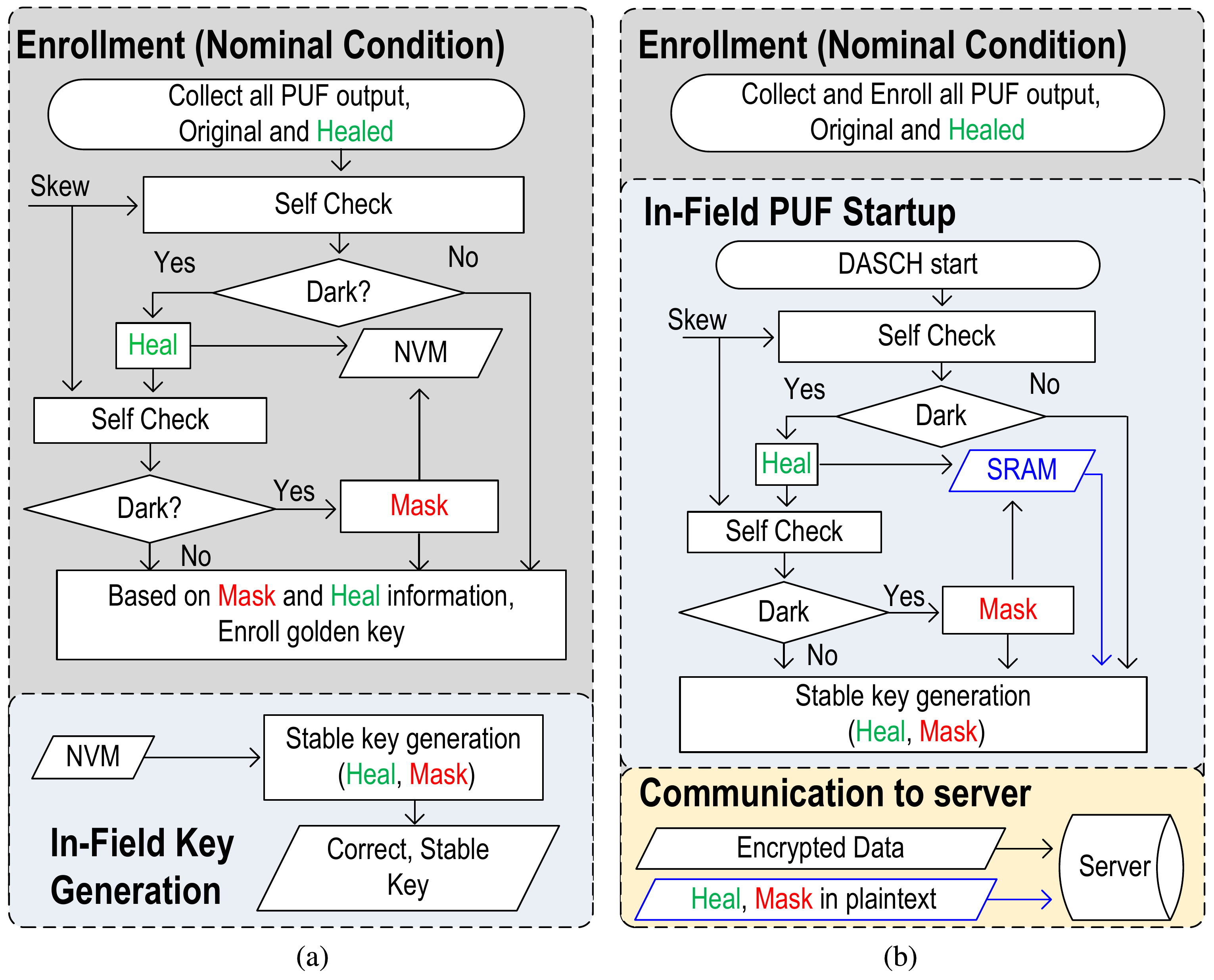}
      \vskip -2ex
      \caption{Operation flows for (a) S-ASCH and (b) D-ASCH.}
      \label{ASCH_D-ASCH_flow}
  \end{figure}
  
Its operation flow is shown in Fig.~\ref{ASCH_D-ASCH_flow}(a). At the start of enrollment, all PUF outputs, original and healed, are collected. During S-ASCH, all PUF cells first go through one round of self-checking. The voltage skew is externally programmed to target a specific BER. After self-checking, the cells that are detected as dark go through a healing process, and the healing information is recorded. All the healed cells go through another round of self-checking, and if it is still detected as dark, it is unable to be healed and therefore masked. The masking information is also stored. Finally, based on the Mask and Heal information gathered from S-ASCH, the previously gathered PUF outputs are processed into a stable PUF key and enrolled into the server as the golden key.

To qualitatively analyze the benefits provided by the S-ASCH system, we categorized the PUF cells into 3 types: the originally stable PUF cells $C_{1}$, the healed stable PUF cells $C_{2}$, and the unstable cells $C_{3}$. In order to output a stable golden key during in-field conditions, $C_{1}$ can be directly used, and $C_{2}$ requires its location written into NVM during enrollment so that system can locate $C_{2}$ and heal it to get the correct and stable key, and $C_{3}$ not only needs an NVM storage for itself, it also needs a redundant stable PUF bit $C_{1}'$, and extra logic to make sure that when the PUF key is generated, $C_{3}$ output will be replaced with $C_{1}'$. So the additional testing cost $T$ required for the PUF bits can be compared as: 
    \begin{equation}
        \begin{split}
            T(C_{3}) &> T(C_{2}) \\
            T(C_{1}) &= 0
        \end{split}
    \end{equation}

As shown in Fig.~\ref{System}, S-ASCH functions by utilizing the components from self-checking and healing and therefore doesn't incur additional design overhead. 
It removes the cost of error correction blocks by achieving an ultra-low BER that meets the typical 1E-6 KER requirements of cryptographic applications, as is shown in section IV.
The proposed dark bit detection method inherently has a better detection accuracy than previous designs, i.e, the number of $C_{3}$ is smaller. Additionally, the healing operation converts a portion of $C_{3}$ to $C_{2}$, further reducing the number of masked cells. 
In conclusion, S-ASCH effectively reduces the design overhead for implementing a 0 BER PUF array both in terms of error correction and NVM storage.

\subsection{Dynamic ASCH (D-ASCH)}
Despite reduced design cost, S-ASCH still requires NVM for healing and masking information storage, which is not ideal in certain IoT applications where the area constraint is extremely tight. Therefore, D-ASCH is conceived, which utilizes the idea of soft-masking to achieve ``0" BER without NVM storage.

Its operation flow is shown in Fig.~\ref{ASCH_D-ASCH_flow}(b). During enrollment, the PUF values of both original and reconfigured cells are collected in the server. D-ASCH operates at every chip power-up in the field. The self-checking and healing operation is the same as S-ASCH, and the generated key is used as the correct and stable key for cryptographic purposes. Instead of writing the healing and masking information to NVM, D-ASCH writes the information to a low-cost SRAM look-up table (LUT). As shown in Fig.~\ref{System}, SRAM-based LUT and a digital stabilization module are added on top of S-ASCH to perform D-ASCH operations. SRAM storage is readily available and cheap to integrate into any chip, so its overhead is negligible compared with NVM implementation. The stabilization module accesses the SRAM for healing and masking information and generates a stable fixed-length key by sequentially reading out the stable PUF cells based on the healing and masking information. When a PUF bit is flagged as stable-after-heal, D-ASCH will heal the PUF cell and read out its value.

\begin{figure}[t]
      \centering
      \includegraphics[scale=0.2]{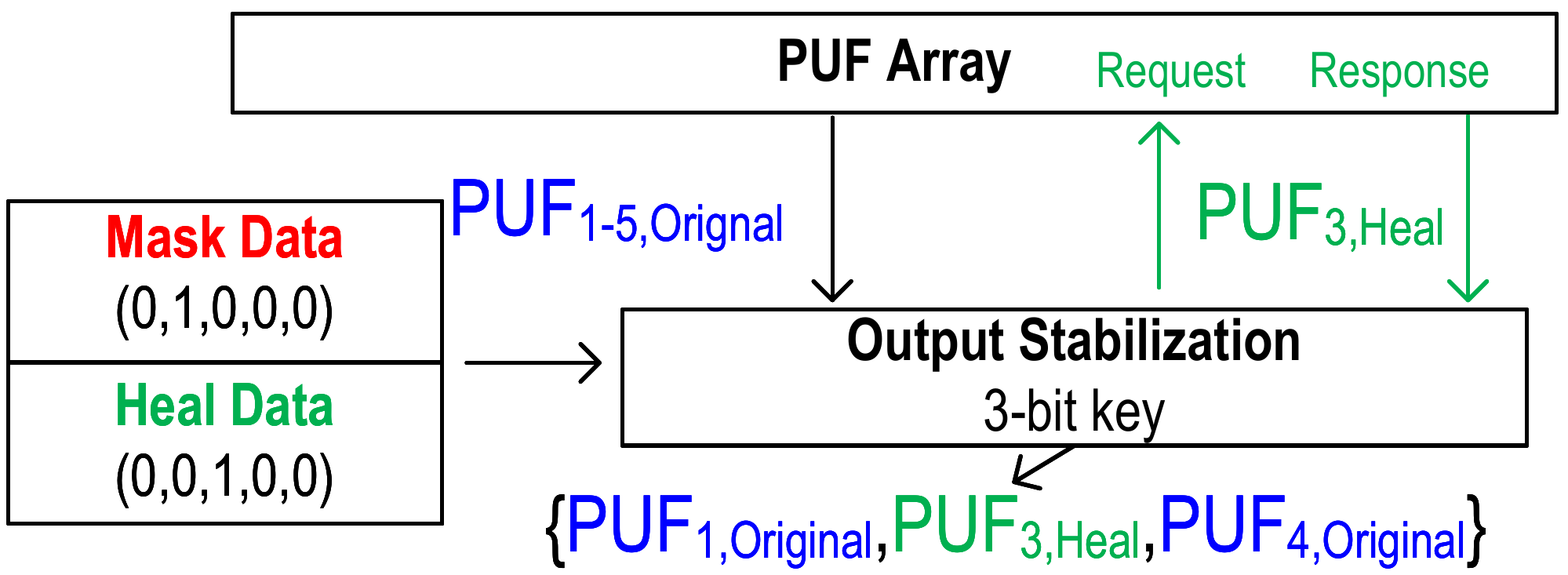}
      \caption{D-ASCH output stabilization example.}
      \label{D-ASCH_example}
  \end{figure}

An example of a 5-bit input, 3-bit output stabilization is shown in Fig.~\ref{D-ASCH_example}. For a row of five PUF cells, $PUF_{2}$ is masked, and $PUF_{3}$ is healed. The stabilization receives the original PUF values and, based on the masking and healing information, omits the $PUF_{2}$ and requests the $PUF_{3, Heal}$. So the final 3-bit key output is: \{$PUF_{1},PUF_{3,Heal},PUF_{4}$\}. Here, redundant PUF cells are still required, but the overhead of implementing more PUF bits using the proposed compact and ultra-low-power PUF cell is insignificant compared with the savings from removing NVM-based masking storage.

The healing and masking information is sent to the server prior to using the PUF keys
because a fresh healing and masking map is generated for every chip start-up and is necessary for servers to properly verify the PUF device. This information can be communicated in plaintext with no concern of eavesdropping because it is random and independent of the actual PUF values.

Similar to S-ASCH, D-ASCH removes the error correction cost for stable key generation. It further eliminates the requirement of NVM for masking storage, which significantly reduces the area overhead for a complete PUF design. 
The downside of this operation is that D-ASCH needs to run the ASCH process for the entire PUF array during every chip power-up. The overhead for the device carrying D-ASCH would be the time, power consumption, and communication effort, all of which are negligible and discussed as follows: 

The amount of time and power to run D-ASCH is insignificant for a chip cold start-up process. It has been stated previously that the entire self-checking operation can be finished within 4ms. We simulated the ASCH operation under nominal conditions (1.2V VDD, 25ºC) in Hspice, and the total energy (including clock generation) for 4ms of ASCH is 1.7$\mu$J.

As for the additional communication effort,
D-ASCH needs to send the masking and healing information in plaintext to the server during communication (Fig.~\ref{ASCH_D-ASCH_flow}b). Assuming an implausible worst case where all the key bits are either masked or healed, this means additional plaintext of a maximum of the key length would need to be communicated along with the encrypted data. In cryptography, the key length is usually trivial compared with the length of information being encrypted. Therefore, additional communication effort is negligible.


\section{measurement results}
\begin{figure}[t]
      \centering
      \includegraphics[width=\columnwidth]{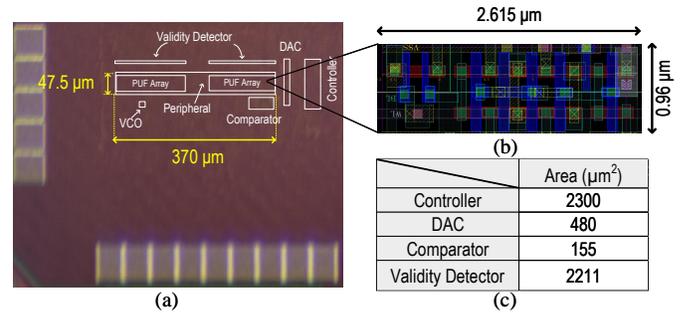}
      \vskip -2ex
      \caption{(a) Chip Micrograph. (b) PUF Cell layout. (c) Area cost of other components for stabilization system.}
      \label{ChipMicrograph}
  \end{figure}

The chip is fabricated using a 65-nm CMOS process. The 128 $\times$ 32 cell array occupies $0.018 mm^2$. The die micrograph, the layout of the PUF cell, and the area overhead of the components needed for S-ASCH are shown in Fig.~\ref{ChipMicrograph}. Each PUF cell measures $0.96 \mu m \times 2.615 \mu m$, or 594 $F^2$. Aside from Validity Detector, which is implemented once per column, the size of all other additional components does not increase with larger PUF array implementation. A voltage-controlled oscillator (VCO) is integrated to provide high-speed clocking for stabilization control and key access. The NVM storage for S-ASCH, SRAM LUT, and stabilization block for D-ASCH shown in Fig.~\ref{System} is not implemented in this prototype.

The nominal condition for the PUF chip is 25 \textdegree C and $1.2V$ supply voltage. Golden keys are collected under nominal conditions by averaging our random noise with many samples. The BER and unstable bit percentage results are measured by comparing separately collected samples under nominal and V/T variations with the golden key.

\subsection{Self Checking and Healing}
\begin{figure}[t]
      \centering
      \includegraphics[width=0.85\columnwidth]{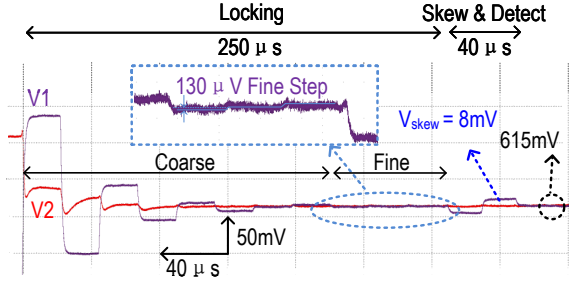}
      \caption{Measured operating waveforms of the self-checking process with a zoomed-in view of fine locking.}
      \label{LockingWaveform_real}
  \end{figure}

The fast and reliable self-checking operation is the foundation of the proposed stabilization system. Supply voltage $V_{1}$ and $V_{2}$ shared by the entire PUF array are output via analog buffer, and the waveform for a self-checking operation is shown in Fig.~\ref{LockingWaveform_real}. The fine step enabled by 4-bit PWM achieves 130 $\mu V$ resolution, which is sufficient to reduce the initial mismatch between two supply voltages to a negligible amount. After applying $-8 mV$ and $8 mV$ voltage skews, the dark bit information is output by the validity detector, and the self-checking process is complete.

The two voltages remain locked after the skewing operations. (see $V_{1}$ and $V_{2}$ settling to $615 mV$ in Fig.~\ref{LockingWaveform_real}). Thus, the coarse-fine locking process only needs to be performed once during a self-checking operation for the entire PUF array. The skew and detection are performed once for every row. The 8-bit binary-searching coarse locking takes eight cycles to complete, and the 4-bit linear-searching fine locking can take from 1 to $2^4=16$ cycles. So in the worst timing case, the coarse-fine locking operation takes 24 cycles to complete. The skewing operation takes two cycles, and for a 32-row PUF array, it takes 64 cycles to complete all the dark bit detection. According to Fig.~\ref{LockingWaveform_real}, the self-checking cycle period is around 20 $\mu$s. Therefore, the maximum timing cost for a self-checking operation is $(24+64)\times20=1760\mu s$.

According to Fig.~\ref{ASCH_D-ASCH_flow}, During either S-ASCH or D-ASCH, the PUF array is checked twice: once for original PUF cells and another for healed ones. Since self-checking is the most time-consuming operation step in the stabilization process, the entire stabilization can be done within $4 ms$. This means almost no timing cost is added to the proposed stabilization system.

\begin{figure}[t]
      \centering
      \includegraphics[width=0.85\columnwidth]{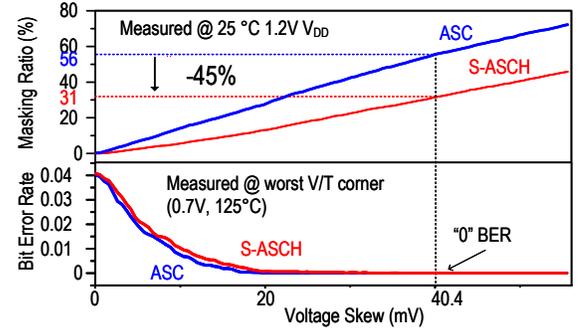}
      \vskip -2ex
      \caption{Masking ratio and resulting BER at different voltage skew, in ASC only and S-ASCH modes.}
      \label{ASCH Direct}
    \end{figure}

To demonstrate the effectiveness of the self-checking and healing process built into the proposed stabilization system, we compare the BER and masking ratio versus the voltage skew at the worst V/T corner of 0.7 V and 125 \textdegree C and show the result of Automatic Self Checking only (ASC), where all the detected dark bits after the first round of self-checking are masked, versus the result of S-ASCH. As shown in Fig.~\ref{ASCH Direct}, the increasing voltage skew represents more aggressive self-checking, leading to an increase in the masking ratio and a decrease in BER. Both ASC and S-ASCH achieve ``0" BER at the same voltage skew, which means the last dark bit found in S-ASCH is not able to be healed. The final masking ratio for the ASC approach is 56\%, which is marginally better than the state-of-the-art design with a similar V/T range~\cite{liu2020373}, without the burden of an additional voltage variation correction step. S-ASCH further reduces the masking ratio to 31\% by healing a portion of the PUF cell, which reduces the PUF redundancy overhead, explained in section III.


\subsection{PUF Stability}

\begin{figure}[t]
      \centering
      \includegraphics[width=0.85\columnwidth]{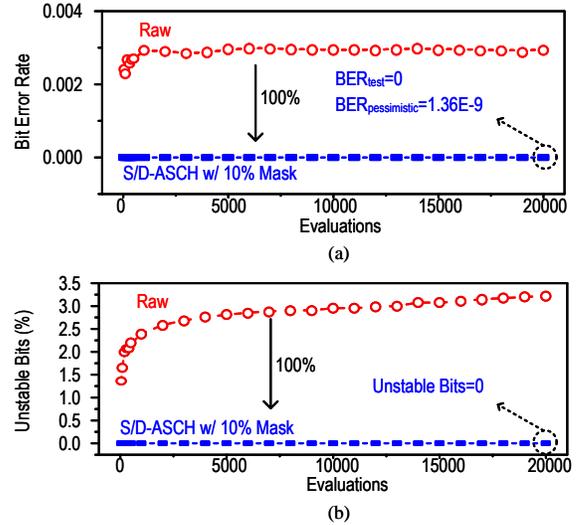}
      \vskip -2ex
      \caption{(a) BER and (b) percentage of unstable bits under nominal conditions. }
      \label{NativeStability}
  \end{figure}

The BER and the proportion of unstable bits of ten chips, measured at the nominal condition before and after stabilization, are depicted in Fig.~\ref{NativeStability}. The two stabilization methods, S-ASCH and D-ASCH, are not distinguished here because the effects are the same under nominal conditions. The BER denotes the rate of bit flips over evaluations compared with the golden value, and its definition is given in Section I. The unstable bit denotes the percentage of ever-flipped bits. The proportion of unstable bits increases as the number of evaluations increases. The raw native BER is at a steady 2.9E-3 without any stabilization methods. Using the proposed stabilization method with a 10\% masking ratio, No error is detected in the testing. At 20000 evaluations, the pessimistic BER according to \eqref{eq:pessimisticBER}, is$$1/(10 \times 4096 \times (1-10\%)\times 20000)=1.36E-9$$ 

The percentage of unstable bits at nominal condition is 3.2\% with 20000 evaluations before stabilization. It is reduced to 0 after 10\% masking.

   \begin{figure}[t]
      \centering
      \includegraphics[width=0.85\columnwidth]{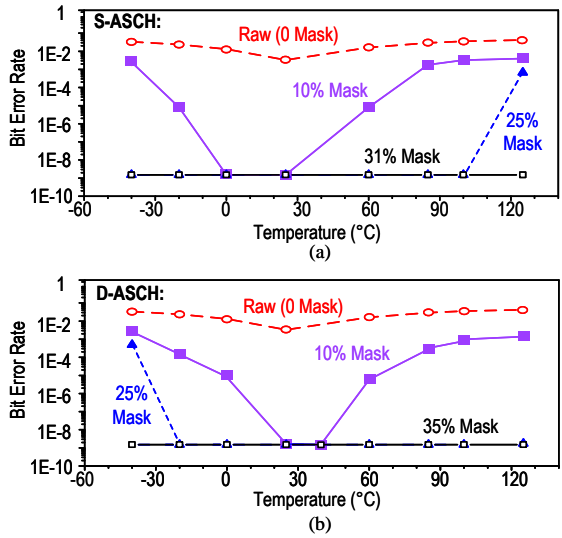}
      \vskip -2ex
      \caption{BER across temperature for raw PUF bits, and after different masking ratios using (a) S-ASCH and (b) D-ASCH. }
      \label{ASCH_D-ASCH_TEMP}
  \end{figure}
  
     \begin{figure}[t]
      \centering
      \includegraphics[width=0.85\columnwidth]{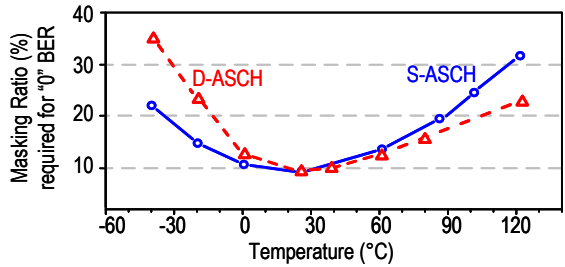}
      \vskip -2ex
      \caption{Masking ratio to achieve ``0" BER for S-ASCH and D-ASCH. }
      \label{ASCH_D-ASCH_MR}
  \end{figure}

The stability of the PUF over environmental variations is evaluated across the automotive temperature range (-45 \textdegree C to 125 \textdegree C) and a voltage range from 0.7V to 1.4V. The golden key is enrolled at a nominal condition. The BER over the temperature sweep is 4.2E-2 before stabilization. S-ASCH and D-ASCH  are separately evaluated.

Fig.~\ref{ASCH_D-ASCH_TEMP}(a) presents the BER across temperature variation where S-ASCH is performed at 25 \textdegree C.
By changing the masking ratio by providing different amounts of voltage skew, ``0" BER can be achieved in the different temperature ranges. At 10\% masking, ``0" BER of \textless 1.36E-9 can be achieved in 0 \textdegree C to 25 \textdegree C. At a more aggressive 31\% masking ratio, ``0" BER of \textless 1.77E-9 is achieved across the entire testing temperature range.

D-ASCH is performed every time during chip start-up, which means it works under various in-field conditions. BER result using D-ASCH is shown in Fig.~\ref{ASCH_D-ASCH_TEMP}(b). At 25\% masking, D-ASCH can reduce the BER to 0 under all temperature conditions higher than enrollment. But 35\% masking is required to deal with the dark bits that become unstable at -40 \textdegree C, achieving ``0" BER of 1.88E-9.

The masking ratio required to achieve ``0" BER for the two stabilization methods is compared under the testing temperature range in Fig.~\ref{ASCH_D-ASCH_MR}. It shows that S-ASCH can reduce the masking ratio more effectively at low temperatures, while D-ASCH performs better at high temperatures. But compared with the huge difference in the storage trade-off of the two stabilization methods, the 10\% masking ratio difference is usually not the deciding factor for which one is more suitable for a certain application. 
Fig.~\ref{ASCH_D-ASCH_MR} can be used as a look-up table when deciding the skew value or masking ratio while targeting a ``0" BER for a specific temperature range. 

\begin{figure}[t]
      \centering
      \includegraphics[scale=0.19]{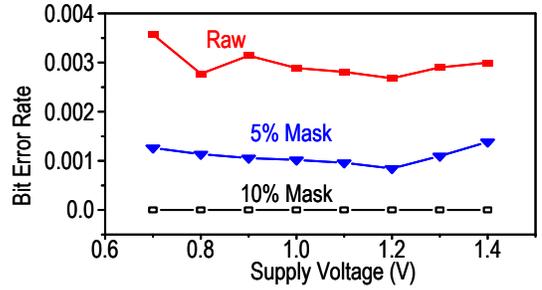}
      \vskip -2ex
      \caption{BER of raw PUF bits, S-ASCH with 5\% masking and with 10\% masking versus supply voltage.}
      \label{Voltage}
  \end{figure}

\begin{figure}[t]
      \centering
      \includegraphics[scale=0.55]{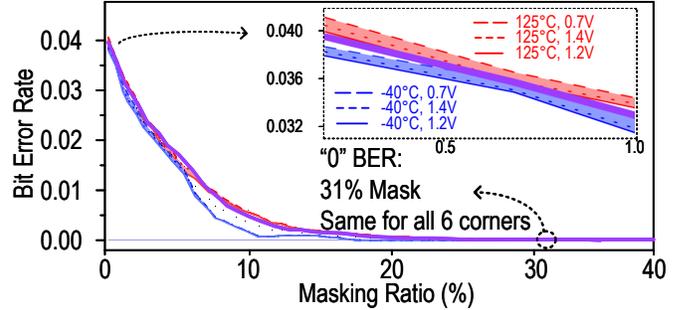}
      \vskip -2ex
      \caption{BER versus masking ratio for S-ASCH at worst temperature and six voltage corners.}
      \label{ASCH CORNER}
  \end{figure}

In addition to temperature variation, we also evaluate the PUF’s resistance to supply voltage variation from 0.7 to 1.4 V, as shown in Fig.~\ref{Voltage}(a). Thanks to the supply regulation based on native transistors, only 0.4\% BER is observed across the voltage range and is reduced to 0 with 10\% masking using S-ASCH. As depicted in Fig.~\ref{ASCH CORNER}(b), the effect of voltage variation is so small compared with that of the temperature variation that at the two extreme temperature corners, voltage variation doesn't change the masking ratio needed to achieve ``0" BER.

     \begin{figure}[t]
      \centering
      \includegraphics[width=0.85\columnwidth]{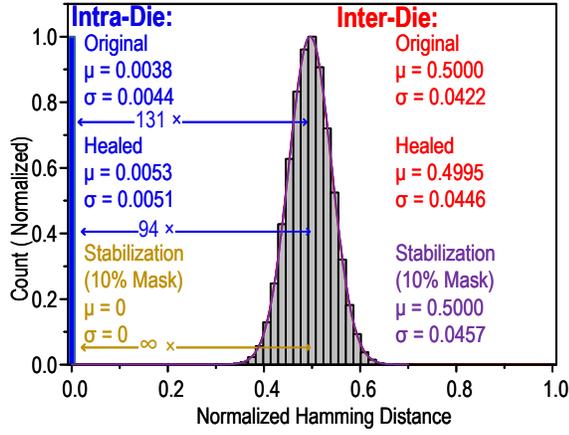}
      \vskip -2ex
      \caption{Normalized Hamming Distances (HD) for original, healed, and using proposed stabilization with 10\% mask, at nominal condition.}
      \label{HD}
  \end{figure}

   \begin{figure}[t]
      \centering
      \includegraphics[width=0.85\columnwidth]{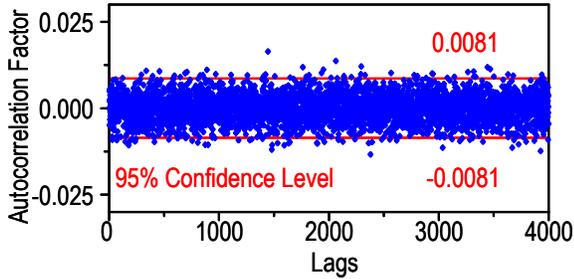}
      \vskip -2ex
      \caption{Autocorrelation tests of 40,960 PUF bits from 10 chips.}
      \label{ACF}
  \end{figure}

\subsection{Uniqueness and Randomness}
  
The inter-die and intra-die hamming distances (HDs) are depicted in Fig.~\ref{HD}. The inter-die HD of the original and healed PUF cells measured over ten chips have almost ideal mean values of 0.5 and 0.4995, respectively. The mean intra-die HD before stabilization is 0.0044 and 0.0051, achieving 131 and 94 times the separation between inter- and intra-die HDs. After stabilization with a 10\% masking ratio (nominal condition, so no distinction between S-ASCH and D-ASCH), the intra-die HD is 0, showing perfect stability and identifiability. The autocorrelation function of 40,960 PUF bits stays within the 95\% white noise confidence level at 0.0081 (see Fig.~\ref{ACF}). The most ideal HD and autocorrelation results validate the uniqueness of the ASCH-PUF.

To further validate the randomness of the PUF, NIST 800-90B~\cite{nistsp80090b} and 800-22~\cite{nistsp80022} randomness tests are performed on 40,960 bits collected from 10 chips. With the limited number of bits, 10 out of 15 subtests in 800-22 are available. NIST-recommended settings were used to run the tests. The PUF bits passed all available subtests in the two suites, showing high-quality randomness as shown in Tables~\ref{NIST90B} and~\ref{NIST22} .

\begin{table}[t]
\caption{\textbf{NIST PUB 800-90B Results}}
\label{NIST90B}
\centering
\setlength{\tabcolsep}{3.6pt}
\renewcommand{\arraystretch}{1.15}
\vskip -1.5ex
\begin{tabular}{|p{70pt}|p{40pt}}
\hline

\multicolumn{1}{|c|}{\textbf{NIST PUB 800-90B (Draft 2)}}
& \multicolumn{1}{c|}{\textbf{Results of 10 chips $\times$ 4096 bits}} 
 
\\\hline

\multicolumn{1}{|c|}{\textbf{IID Permutation}}
& \multicolumn{1}{c|}{{PASS}} 
\\\hline

\multicolumn{1}{|c|}{\textbf{$\chi^2$ Independence}}
& \multicolumn{1}{c|}{{PASS (Score=2027.26, dof=2047)}} 
\\\hline

\multicolumn{1}{|c|}{\textbf{$\chi^2$ Goodness-of-fit}}
& \multicolumn{1}{c|}{{PASS (Score=6.63718, dof=9)}} 
\\\hline

\multicolumn{1}{|c|}{\textbf{LRS Test}}
& \multicolumn{1}{c|}{{PASS (Pr=0.970268)}} 
\\\hline

\multicolumn{1}{|c|}{\textbf{Min Entropy}}
& \multicolumn{1}{c|}{{0.977}} 
\\\hline

\end{tabular}
\end{table}

\begin{table}[t]
\caption{\textbf{NIST PUB 800-22 Results}}
\label{NIST22}
\centering
\setlength{\tabcolsep}{3.6pt}
\renewcommand{\arraystretch}{1.15}
\vskip -1.5ex
\begin{tabular}{|p{70pt}|p{40pt}|p{40pt}}
\hline
\multicolumn{1}{|c|}{\textbf{NIST Pub 800-22(rev. 1a, 2010)}}
& \multicolumn{1}{c|}{\textbf{$\chi^2$ of p-value}} 
& \multicolumn{1}{c|}{\textbf{Success Proportion}} 
\\\hline

\multicolumn{1}{|c|}{\textbf{Frequency}}
& \multicolumn{1}{c|}{0.876} 
& \multicolumn{1}{c|}{116/120} 

\\\hline
\multicolumn{1}{|c|}{\textbf{Block Frequency}}
& \multicolumn{1}{c|}{0.048} 
& \multicolumn{1}{c|}{118/120} 
\\\hline
\multicolumn{1}{|c|}{\textbf{Cumulative Sum-1}}
& \multicolumn{1}{c|}{0.941} 
& \multicolumn{1}{c|}{116/120} 
\\\hline
\multicolumn{1}{|c|}{\textbf{Cumulativ Sum-2}}
& \multicolumn{1}{c|}{0.723} 
& \multicolumn{1}{c|}{117/120} 
\\\hline
\multicolumn{1}{|c|}{\textbf{Runs}}
& \multicolumn{1}{c|}{0.534} 
& \multicolumn{1}{c|}{119/120} 
\\\hline
\multicolumn{1}{|c|}{\textbf{Longest Runs}}
& \multicolumn{1}{c|}{0.223} 
& \multicolumn{1}{c|}{120/120} 
\\\hline
\multicolumn{1}{|c|}{\textbf{FFT}}
& \multicolumn{1}{c|}{0.804} 
& \multicolumn{1}{c|}{58/60} 
\\\hline
\multicolumn{1}{|c|}{\textbf{Serial-1}}
& \multicolumn{1}{c|}{0.299} 
& \multicolumn{1}{c|}{118/120} 
\\\hline
\multicolumn{1}{|c|}{\textbf{Serial-2}}
& \multicolumn{1}{c|}{0.007} 
& \multicolumn{1}{c|}{118/120} 
\\\hline
\multicolumn{1}{|c|}{\textbf{Approximate Entropy}}
& \multicolumn{1}{c|}{0.264} 
& \multicolumn{1}{c|}{118/120} 
\\\hline
\multicolumn{1}{|c|}{\textbf{Non Overlapping Template}}
& \multicolumn{1}{c|}{{PASS}} 
& \multicolumn{1}{c|}{{PASS}} 
\\\hline

\end{tabular}
\end{table}

\subsection{Aging}

\begin{figure}[t]
      \centering
      \includegraphics[width=0.85\columnwidth]{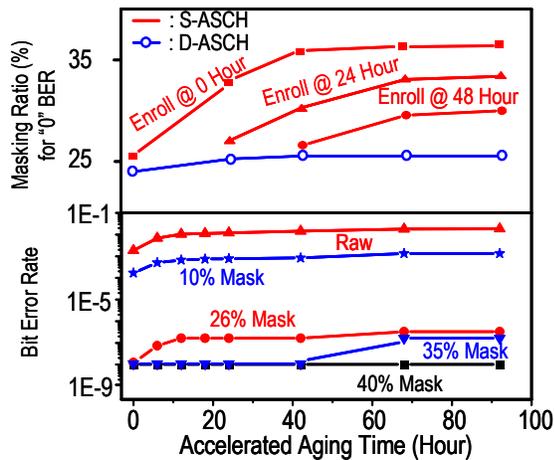}
      \vskip -3ex
      \caption{(bottom) Aging effects on BER with different masking ratios using S-ASCH, and (top) the masking ratio required for ``0" BER using S-ASCH enrollment at different time points (red lines) and using D-ASCH (blue line).}
      \label{aging}
  \end{figure}

It is well known that aging degrades PUF stability or completely flip bits. The main sources of aging effects in PUF are NBTI and HCI~\cite{liu202136,taneja2018fully,li2016ultra,yang201514,stanzione2011cmos}. In order to evaluate the aging impacts, accelerated aging is applied to 2 chips, 4096 PUF bits per chip, by stressing the PUF at 150 \textdegree C and $1.4 V$ supply voltage. The golden key and S-ASCH configuration were obtained at nominal conditions. To evaluate the performance at an extreme scenario using either ASCH system, D-ASCH configurations and the PUF values are collected at 0.7V, 125 \textdegree C.
Measurements were obtained every 6 hours during the first day and every 24 hours after that. A stressing of 96 hours in total was applied, resulting in equivalent effects of several years’ aging under nominal conditions. For S-ASCH, the measured aging-induced BER change with different masking ratios and the masking ratio required to achieve ``0" BER for enrollment at different time points is shown in Fig.~\ref{aging}. The required masking ratio to maintain the same BER increases with aging time, but if the enrollment is run after some time of accelerated aging, the masking ratio will not have such an aggressive increase, as is depicted by the slope decrease for enrollment at 0 hours. This shows that S-ASCH benefits from having a burn-in process prior to enrollment.

We also show the aging result for D-ASCH with enrollment performed at the start of the experiment (blue line in Fig.~\ref{aging} top). The masking ratio to achieve ``0" BER for D-ASCH at the start of aging is 24\% and maintained below 26\% throughout the aging experiment. Compared to the S-ASCH result enrolled at hour 0, the performance of D-ASCH is better than S-ASCH when aging is considered.

This is because S-ASCH needs to increase its detection range ($V_{skew}$) to "predict" the PUF bits that become unstable after aging, which reduces its detection accuracy (shown in Fig.~\ref{Dsim}b). On the other hand, the detection range of D-ASCH doesn’t need to change because this scheme re-evaluates the PUF cells at every chip start-up, and any currently unstable cells will be identified by the default voltage skew. The slight increase in the masking ratio is only caused by the increase of unstable bits from aging.

 \begin{figure}[t]
      \centering
      \includegraphics[width=0.85\columnwidth]{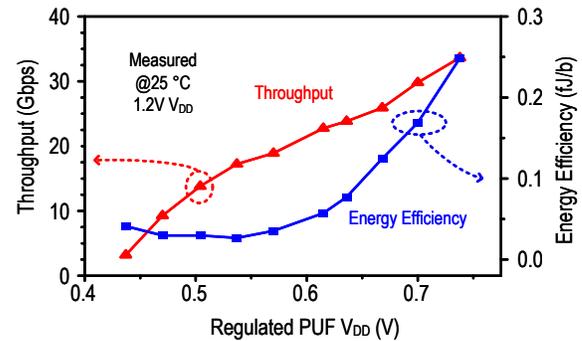}
      \vskip -2ex
      \caption{Throughput and energy efficiency versus regulated PUF $V_{DD}$ at nominal condition.}
      \label{Throughput}
  \end{figure}
  
\subsection{Throughput and Energy Efficiency}
The design reaches $22.75 Gb/s$ readout throughput with $615 mV$ PUF cell $V_{DD}$, thanks to the SRAM-style PUF array and 128-bit parallel readout. The subthreshold PUF consumes merely $0.056 fJ$ per bit of core energy. The measured throughput and energy efficiency at various regulated PUF $V_{DD}$ are shown in Fig.~\ref{Throughput}. The native transistor-based voltage regulation suppresses the supply voltage-induced influence on current. 
A comprehensive comparison with state-of-the-art PUF designs is provided in Table~\ref{comp}.

\begin{table*}[t]
\caption{\textbf{Comparison Table with State-of-the-Art PUF Designs}}
\centering
\setlength{\tabcolsep}{3.6pt}
\renewcommand{\arraystretch}{1.2}
\vskip -1.5ex
\begin{tabular}{|p{35pt}|p{40pt}|p{45pt}|p{45pt}|p{45pt}|p{45pt}|p{45pt}|p{35pt}|p{35pt}|p{35pt}|p{35pt}|}
\hline
\multicolumn{2}{|c|}{} &
\multicolumn{2}{c|}{\textbf{This Work}} & 
\parbox[c][0.8cm]{45pt}{\centering{JSSC 21\\
\cite{liu5VHybridSRAM2021}}}&
\parbox[c][0.8cm]{45pt} {\centering{JSSC 20 \\
\cite{liu2020373}}}&
\parbox[c][0.8cm]{45pt} {\centering{TCAS 20 \\
\cite{shifman2020sram}}}&
\parbox[c][0.8cm]{35pt} {\centering{JSSC 22\\
\cite{park2022ber}}} &
\parbox[c][0.8cm]{35pt} {\centering{ISSCC 20 \\
\cite{choi2020physically}}} &
\parbox[c][0.8cm]{35pt} {\centering{JSSC 18 \\
\cite{taneja2018fully}}} & 
\parbox[c][0.8cm]{35pt} {\centering{JSSC 22 \\
\cite{satpathy20174}}} 
\\\hline

\multicolumn{2}{|c|}{\textbf{Technology (nm)}} & 
\multicolumn{2}{c|}{\textbf{65}}& 
\multicolumn{1}{c|}{130}& 
\multicolumn{1}{c|}{130}&
\multicolumn{1}{c|}{65}&
\multicolumn{1}{c|}{40}&
\multicolumn{1}{c|}{28}&
\multicolumn{1}{c|}{40}&
\multicolumn{1}{c|}{14}
\\\hline

\multicolumn{2}{|c|}{\textbf{PUF Cell Area/Bit (F$^2$)}} 
& \multicolumn{2}{c|}{\textbf{594}}
& \multicolumn{1}{c|}{491}
& \multicolumn{1}{c|}{373}
& \multicolumn{1}{c|}{3001} 
& \multicolumn{1}{c|}{21675} 
& \multicolumn{1}{c|}{5057}
& \multicolumn{1}{c|}{3643}
& \multicolumn{1}{c|}{9388}
\\\hline

\multicolumn{2}{|c|}{\textbf{Native Unstable Bits}} 
& \multicolumn{2}{c|}{\textbf{3.2\%}}
& \multicolumn{1}{c|}{2.71\%}
& \multicolumn{1}{c|}{2.14\%}
& \multicolumn{1}{c|}{19.6\%$^c$}
& \multicolumn{1}{c|}{0.39\%}
& \multicolumn{1}{c|}{-}
& \multicolumn{1}{c|}{2.55\%}
& \multicolumn{1}{c|}{26.37\%$^c$}
\\\hline

\multicolumn{2}{|c|}{\textbf{Native BER}} 
& \multicolumn{2}{c|}{\textbf{2.90E-03}}
& \multicolumn{1}{c|}{2.90E-03}
& \multicolumn{1}{c|}{2.10E-03}
& \multicolumn{1}{c|}{2.24E-02}
& \multicolumn{1}{c|}{2.7E-04}
& \multicolumn{1}{c|}{-}
& \multicolumn{1}{c|}{8.10E-03}
& \multicolumn{1}{c|}{-}
\\\hline

\multirow{2}{*}{\textbf{\shortstack{Test\\Condition}}} &
\multicolumn{1}{c|}{\textbf{Temp.(\textdegree C)}} 
& \multicolumn{2}{c|}{\textbf{-40-125}}
& \multicolumn{1}{c|}{-40-120}
& \multicolumn{1}{c|}{-40-120}
& \multicolumn{1}{c|}{-10-85}
& \multicolumn{1}{c|}{-40-125}
& \multicolumn{1}{c|}{-40-150}
& \multicolumn{1}{c|}{-40-125}
& \multicolumn{1}{c|}{25-110}
\\\cline{2-11}

\multicolumn{1}{|c|}{} 
&\multicolumn{1}{c|}{\textbf{Supply (V)}} 
& \multicolumn{2}{c|}{\textbf{0.7-1.4}}
& \multicolumn{1}{c|}{0.5-0.7}
& \multicolumn{1}{c|}{0.8-1.4}
& \multicolumn{1}{c|}{0.8-1.2} 
& \multicolumn{1}{c|}{0.7-1.4} 
& \multicolumn{1}{c|}{0.81-0.99}
& \multicolumn{1}{c|}{0.8-1.0}
& \multicolumn{1}{c|}{0.7-1.0.75}
\\\hline

\multicolumn{2}{|c|}{\textbf{Stabilizing Technique$^a$}}&
\parbox[c][1.5cm]{45pt}{\centering
{\textbf{S-ASCH}}}&
\parbox[c][0.8cm]{45pt}{\centering
{\textbf{D-ASCH}}}&
\parbox[c][0.8cm]{45pt}{\centering
{HCI Burn-In}}&
\parbox[c][0.8cm]{45pt} {\centering{VSS Bias,\\Voltage Overdrive, \\ Masking}}&
\parbox[c][0.8cm]{45pt} {\centering{Capacitive Tilt, \\ Masking}}&
\parbox[c][0.8cm]{35pt} {\centering{Masking,\\ TMV}} &
\parbox[c][0.8cm]{35pt} {\centering{Masking, \\ SMV}} &
\parbox[c][0.8cm]{35pt} {\centering{Hysteresis, \\ T Comp.}} &
\parbox[c][0.8cm]{35pt} {\centering{Masking, \\ Burn-in,\\ TMV}} 
\\\hline

\multicolumn{2}{|c|}{\textbf{NVM requirement$^b$}} 
& \multicolumn{1}{c|}{\textbf{Yes}}
& \multicolumn{1}{c|}{\textbf{No}}
& \multicolumn{1}{c|}{No}
& \multicolumn{1}{c|}{Yes}
& \multicolumn{1}{c|}{Yes}
& \multicolumn{1}{c|}{Yes}
& \multicolumn{1}{c|}{Yes}
& \multicolumn{1}{c|}{No}
& \multicolumn{1}{c|}{Yes}
\\\hline

\multirow{5}{*}{\textbf{\shortstack{Worst\\Condition}}} &
\parbox[c][0.8cm]{41pt}{\centering
{\textbf{BER before\\ stabilization}}} &
\multicolumn{2}{c|}{\textbf{4.20E-02}}
& \multicolumn{1}{c|}{6.74E-02}
& \multicolumn{1}{c|}{5.80E-02} 
& \multicolumn{1}{c|}{4.00E-02$^d$} 
& \multicolumn{1}{c|}{-}
& \multicolumn{1}{c|}{1.05E-01}
& \multicolumn{1}{c|}{3.20E-02}
& \multicolumn{1}{c|}{5.76E-02}
\\\cline{2-11}

\multicolumn{1}{|c|}{}&
\parbox[c][0.8cm]{41pt}{\centering
{\textbf{BER after\\ stabilization}}} &
\parbox[c][0.8cm]{45pt}{\centering
{\textbf{0\\(\textless1.77E-09) }}}&
\parbox[c][0.8cm]{45pt}{\centering
{\textbf{0\\ (\textless1.88E-09)}}}&
\parbox[c][0.8cm]{45pt} {\centering
{0\\ (\textless4.0E-07$^e$)}}&
\parbox[c][0.8cm]{45pt} {\centering
{0\\ (\textless5.99E-07)}}&
\parbox[c][0.8cm]{40pt} {\centering
{0\\ (\textless1.0E-09)}}&

\multicolumn{1}{c|}{1.9E-05}  
& \multicolumn{1}{c|}{9.70E-03}
& \multicolumn{1}{c|}{-}
& \multicolumn{1}{c|}{1.42E-02}
\\\cline{2-11}

\multicolumn{1}{|c|}{}&
\parbox[c][0.8cm]{41pt}{\centering
{\textbf{Masking\\Ratio}}}
& \multicolumn{1}{c|}{\textbf{31\%}} 
& \multicolumn{1}{c|}{\textbf{35\%}} 
& \multicolumn{1}{c|}{-}
& \multicolumn{1}{c|}{67\%}
& \multicolumn{1}{c|}{59\%}
& \multicolumn{1}{c|}{3.64\%}
& \multicolumn{1}{c|}{25\%}
& \multicolumn{1}{c|}{-}
& \multicolumn{1}{c|}{20\%}
\\\hline

\multicolumn{2}{|c|}{\textbf{Bit Rate (Mb/s)}}  
& \multicolumn{2}{c|}{\textbf{22750}} 
& \multicolumn{1}{c|}{56}
& \multicolumn{1}{c|}{-} 
& \multicolumn{1}{c|}{-}
& \multicolumn{1}{c|}{-}
& \multicolumn{1}{c|}{128}
& \multicolumn{1}{c|}{24000}
& \multicolumn{1}{c|}{-}
\\\hline

\multicolumn{2}{|c|}{\textbf{Core Energy (fJ/bit)}}  
& \multicolumn{2}{c|}{\textbf{0.057}} 
& \multicolumn{1}{c|}{128} 
& \multicolumn{1}{c|}{16}
& \multicolumn{1}{c|}{15.39}
& \multicolumn{1}{c|}{39}
& \multicolumn{1}{c|}{-}
& \multicolumn{1}{c|}{1.02}
& \multicolumn{1}{c|}{4}
\\\hline

\multicolumn{2}{l}{a: Not including ECC}
&\multicolumn{6}{l}{b: Including NVM storage for masking, ECC, and other stabilization methods}
&\multicolumn{2}{l}{c: Under V/T Variation}
\\
\multicolumn{4}{l}{d: Averaged across multiple V/T conditions}
&\multicolumn{3}{l}{e: Calculated from 5-chip data}

\end{tabular}

\label{comp}
\end{table*}

\section{Conclusion}
In conclusion, this article presents two stabilization methods, S-ASCH and D-ASCH, integrated with a subthreshold inverter-based PUF array. Utilizing a novel, fast, and high-precision searching method of emulating temperature variation by skewing the first-stage inverter voltage and the reconfiguration capability of the PUF cell, S-ASCH effectively reduces the BER to the order of 1E-9 at the enrollment stage across a wide range of voltage and temperature corners, which reduces the cost of error correction, and in particular removes the need of error correction for a 128-bit key with KER requirement above 3E-7.
The required 31\% masking ratio is less than half of the state-of-the-art design, representing a significant reduction of design overhead of NVM storage implementation and the PUF cell redundancy. 
D-ASCH achieves ``0'' BER with a slightly larger masking ratio at the worst scenario, but the major cost of masking on the edge device, the large NVM storage, is eliminated by using larger storage on the server. The two modes of operation can be chosen based on target applications considering operating conditions, costs, and stability targets. 
Our 65nm PUF occupies 594 F$^2$ area for each bit. The measured responses from 10 chips pass all applicable NIST 800-22 and 800-90B randomness tests. S-ASCH and D-ASCH both achieve ``0" BER with the least amount of design overhead compared with other state-of-the-art designs, making them suitable to provide low-cost key generation and storage for a wide range of applications. Finally, the ASCH-PUF prototype achieves state-of-the-art throughput of 11.4 Gbps with 0.057 fJ/b core energy efficiency at 1.2V, 25°C conditions.


%

\appendices



\ifCLASSOPTIONcaptionsoff
  \newpage
\fi



\bibliography{JSSC.bib}

\begin{thebibliography}{10}

\bibitem{yang2017security}
K.~Yang, D.~Blaauw, and D.~Sylvester, ``Hardware designs for security in
  ultra-low-power iot systems: An overview and survey,'' {\em IEEE Micro},
  vol.~37, no.~6, pp.~72--89, 2017.

\bibitem{taneja2021puf}
S.~Taneja and M.~Alioto, ``{PUF} architecture with run-time adaptation for
  resilient and energy-efficient key generation via sensor fusion,'' {\em IEEE
  Journal of Solid-State Circuits}, 2021.

\bibitem{liu5VHybridSRAM2021}
K.~Liu, X.~Chen, H.~Pu, and H.~Shinohara, ``A 0.5-{{V Hybrid SRAM Physically
  Unclonable Function Using Hot Carrier Injection Burn-In}} for {{Stability
  Reinforcement}},'' {\em IEEE Journal of Solid-State Circuits}, vol.~56,
  pp.~2193--2204, July 2021.

\bibitem{li2019self}
D.~Li and K.~Yang, ``{A self-regulated and reconfigurable CMOS physically
  unclonable function featuring zero-overhead stabilization},'' {\em IEEE
  Journal of Solid-State Circuits}, vol.~55, no.~1, pp.~98--107, 2019.

\bibitem{liu2020373}
K.~Liu, Y.~Min, X.~Yang, H.~Sun, and H.~Shinohara, ``A 373-{$F^2$}
  0.21\%-native-{BER EE SRAM} physically unclonable function with {2-D}
  power-gated bit cells and {VSS} bias-based dark-bit detection,'' {\em IEEE
  Journal of Solid-State Circuits}, vol.~55, no.~6, pp.~1719--1732, 2020.

\bibitem{choi2020physically}
Y.~Choi, B.~Karpinskyy, K.-M. Ahn, Y.~Kim, S.~Kwon, J.~Park, Y.~Lee, and
  M.~Noh, ``Physically unclonable function in 28nm fdsoi technology achieving
  high reliability for aec-q 100 grade 1 and iso 26262 asil-b,'' in {\em 2020
  IEEE International Solid-State Circuits Conference-(ISSCC)}, pp.~426--428,
  IEEE, 2020.

\bibitem{taneja2018fully}
S.~Taneja, A.~B. Alvarez, and M.~Alioto, ``{Fully synthesizable PUF featuring
  hysteresis and temperature compensation for 3.2\% native BER and 1.02 fJ/b in
  40 nm},'' {\em IEEE Journal of Solid-State Circuits}, vol.~53, no.~10,
  pp.~2828--2839, 2018.

\bibitem{karpinskyy20168}
B.~Karpinskyy, Y.~Lee, Y.~Choi, Y.~Kim, M.~Noh, and S.~Lee, ``{Physically
  unclonable function for secure key generation with a key error rate of 2E-38
  in 45nm smart-card chips},'' in {\em 2016 IEEE International Solid-State
  Circuits Conference (ISSCC)}, pp.~158--160, IEEE, 2016.

\bibitem{lee2018445f}
J.~Lee, D.~Lee, Y.~Lee, and Y.~Lee, ``{A $445F^2$ leakage-based physically
  unclonable function with lossless stabilization through remapping for IoT
  security},'' in {\em 2018 IEEE International Solid-State Circuits
  Conference-(ISSCC)}, pp.~132--134, IEEE, 2018.

\bibitem{satpathy20174}
S.~Satpathy, S.~K. Mathew, V.~Suresh, M.~A. Anders, H.~Kaul, A.~Agarwal, S.~K.
  Hsu, G.~Chen, R.~K. Krishnamurthy, and V.~K. De, ``{A 4-fJ/b delay-hardened
  physically unclonable function circuit with selective bit destabilization in
  14-nm trigate CMOS},'' {\em IEEE Journal of Solid-State Circuits}, vol.~52,
  no.~4, pp.~940--949, 2017.

\bibitem{li2016ultra}
J.~Li and M.~Seok, ``{Ultra-compact and robust physically unclonable function
  based on voltage-compensated proportional-to-absolute-temperature voltage
  generators},'' {\em IEEE Journal of Solid-State Circuits}, vol.~51, no.~9,
  pp.~2192--2202, 2016.

\bibitem{yang20178}
K.~Yang, Q.~Dong, D.~Blaauw, and D.~Sylvester, ``{A 553$F^2$ 2-transistor
  amplifier-based Physically Unclonable Function (PUF) with 1.67\% native
  instability},'' in {\em 2017 IEEE International Solid-State Circuits
  Conference (ISSCC)}, pp.~146--147, IEEE, 2017.

\bibitem{park2022ber}
J.~Park, B.~Kim, and J.-Y. Sim, ``A {{BER-Suppressed PUF With}} an
  {{Amplification}} of {{Process Mismatch Effect}} in an {{Oscillator Collapse
  Topology}},'' {\em IEEE Journal of Solid-State Circuits}, vol.~57,
  pp.~2208--2219, July 2022.

\bibitem{mathew201416}
S.~K. Mathew, S.~K. Satpathy, M.~A. Anders, H.~Kaul, S.~K. Hsu, A.~Agarwal,
  G.~K. Chen, R.~J. Parker, R.~K. Krishnamurthy, and V.~De, ``{A 0.19 pJ/b
  PVT-variation-tolerant hybrid physically unclonable function circuit for
  100\% stable secure key generation in 22nm CMOS},'' in {\em 2014 IEEE
  International Solid-State Circuits Conference Digest of Technical Papers
  (ISSCC)}, pp.~278--279, IEEE, 2014.

\bibitem{basak2013reconfigurable}
A.~Basak, S.~Paul, J.~Park, J.~Park, and S.~Bhunia, ``{Reconfigurable ECC for
  adaptive protection of memory},'' in {\em 2013 IEEE 56th International
  Midwest Symposium on Circuits and Systems (MWSCAS)}, pp.~1085--1088, IEEE,
  2013.

\bibitem{shifman2020sram}
Y.~Shifman, A.~Miller, O.~Keren, Y.~Weizman, and J.~Shor, ``{An SRAM-based PUF
  with a capacitive digital preselection for a 1E-9 key error probability},''
  {\em IEEE Transactions on Circuits and Systems I: Regular Papers}, vol.~67,
  no.~12, pp.~4855--4868, 2020.

\bibitem{satpathy201413fj}
S.~Satpathy, S.~Mathew, J.~Li, P.~Koeberl, M.~Anders, H.~Kaul, G.~Chen,
  A.~Agarwal, S.~Hsu, and R.~Krishnamurthy, ``{13fJ/bit probing-resilient 250K
  PUF array with soft darkbit masking for 1.94\% bit-error in 22nm tri-gate
  CMOS},'' in {\em ESSCIRC 2014-40th European Solid State Circuits Conference
  (ESSCIRC)}, pp.~239--242, IEEE, 2014.

\bibitem{lee2021samsung}
Y.~Lee, B.~Karpinskyy, Y.~Choi, K.-M. Ahn, Y.~Kim, J.~Park, S.~Noh, J.~Kang,
  J.~Shin, J.~Park, {\em et~al.}, ``{Samsung Physically Unclonable Function
  (SAMPUF$^{TM}$) and its integration with Samsung Security System},'' in {\em
  2021 IEEE Custom Integrated Circuits Conference (CICC)}, pp.~1--7, IEEE,
  2021.

\bibitem{he202136}
Y.~He, D.~Li, Z.~Yu, and K.~Yang, ``An automatic self-checking and healing
  {Physically Unclonable Function (PUF)} with \textless $3\times10^{-8}$ bit
  error rate,'' in {\em 2021 IEEE International Solid-State Circuits Conference
  (ISSCC)}, vol.~64, pp.~506--508, IEEE, 2021.

\bibitem{nistsp80090b}
NIST, ``{NIST SP 800-90B: Entropy sources used for random bit generation},''
  [Online]. Available:
  https://csrc.nist.gov/publications/detail/sp/800-90b/final.

\bibitem{nistsp80022}
NIST, ``{NIST SP 800-22}: Documentation and software random bit generation,''
  [Online]. Available:
  https://csrc.nist.gov/projects/random-bit-generation/documentation-and-software.

\bibitem{liu202136}
K.~Liu, Z.~Fu, G.~Li, H.~Pu, Z.~Guan, X.~Wang, X.~Chen, and H.~Shinohara, ``A
  modeling attack resilient strong {PUF} with feedback-{SPN} structure having
  \textless 0.73\% bit error rate through in-cell hot-carrier injection
  burn-in,'' in {\em 2021 IEEE International Solid-State Circuits Conference
  (ISSCC)}, vol.~64, pp.~502--504, IEEE, 2021.

\bibitem{yang201514}
T.~Yang, D.~Kim, P.~R. Kinget, and M.~Seok, ``{In-situ techniques for in-field
  sensing of NBTI degradation in an SRAM register file},'' in {\em 2015 IEEE
  International Solid-State Circuits Conference-(ISSCC) Digest of Technical
  Papers}, pp.~1--3, IEEE, 2015.

\bibitem{stanzione2011cmos}
S.~Stanzione, D.~Puntin, and G.~Iannaccone, ``{CMOS silicon physical unclonable
  functions based on intrinsic process variability},'' {\em IEEE Journal of
  Solid-State Circuits}, vol.~46, no.~6, pp.~1456--1463, 2011.

\end{thebibliography}
\bibliographystyle{ieeetr}
%



%

\begin{IEEEbiography}[{\includegraphics[width=1in,height=1.25in,clip,keepaspectratio]{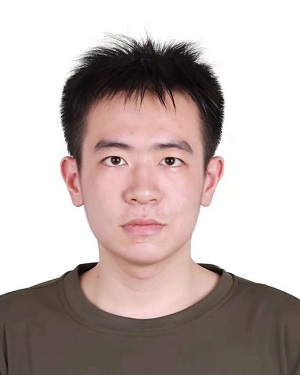}}]{Yan He} received the B.S degree in electronic science and technology from the Zhejiang University, Hangzhou, China, in 2018. He is currently pursuing the Ph.D. degree in electrical and computer engineering with the Rice University, Houston, TX, USA.

His current research interests include analog and mixed-signal integrated circuits design for power management and hardware security. He received the Best Paper Award at the 2021 IEEE Custom Integrated Circuits Conference (CICC). He is the recipient of the 2021-2022 IEEE Solid-State Circuits Society (SSCS) Predoctoral Achievement Award.

\end{IEEEbiography}

\begin{IEEEbiography}[{\includegraphics[width=1in,height=1.25in,clip,keepaspectratio]{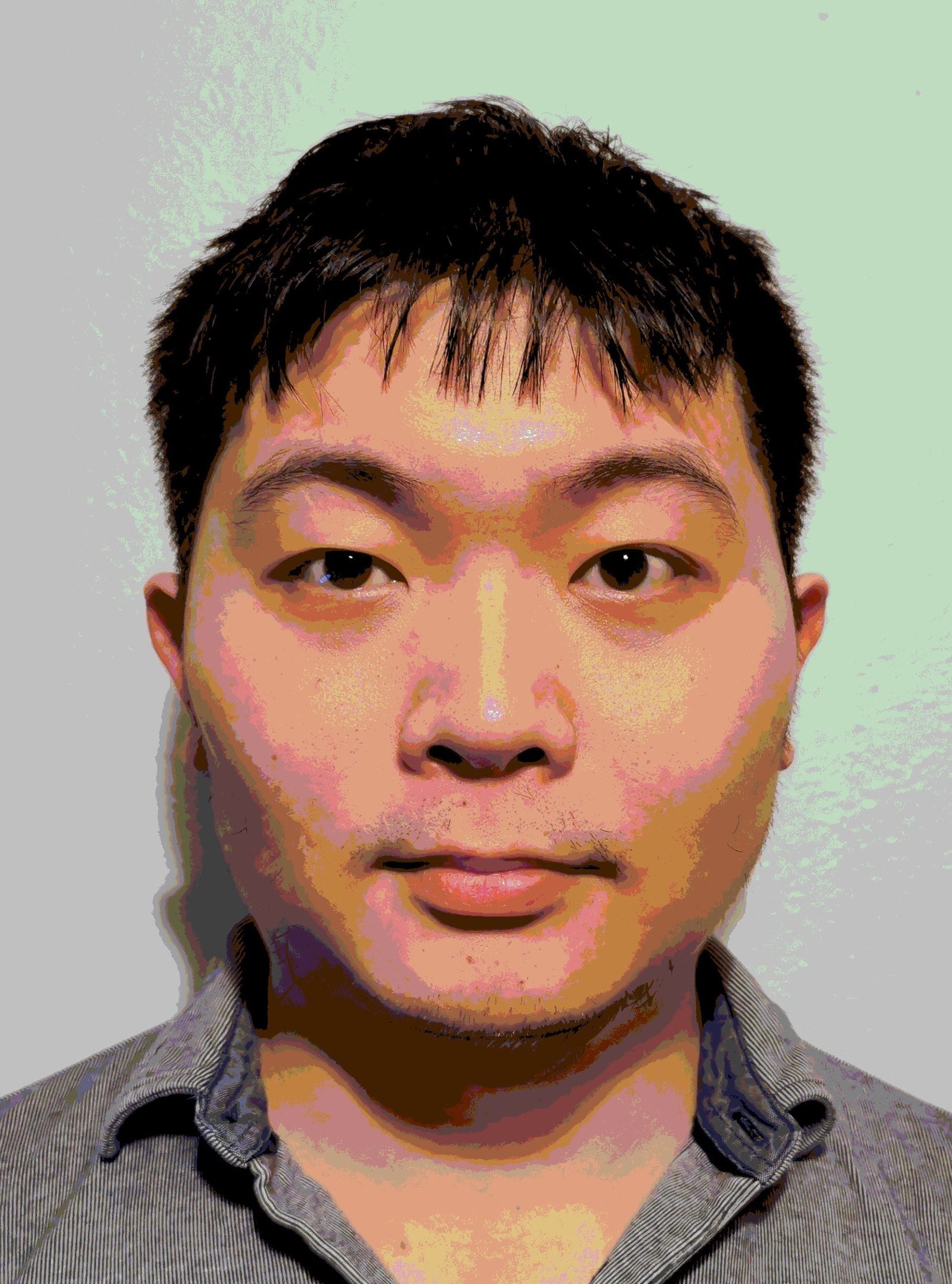}}]{Dai Li} received the B.S. and M.S. degrees in electronic engineering from Tsinghua University, Beijing, China, in 2010 and 2013, respectively, and the PhD degree in electrical and computer engineering from Rice University, Houston, TX, USA,in 2021. He is now at Google.

His research interests include very large-scale integration (VLSI) circuits, hardware security, mixed-signal integrated circuits, and low-power circuits.
\end{IEEEbiography}

\begin{IEEEbiography}[{\includegraphics[width=1in,height=1.25in,clip,keepaspectratio]{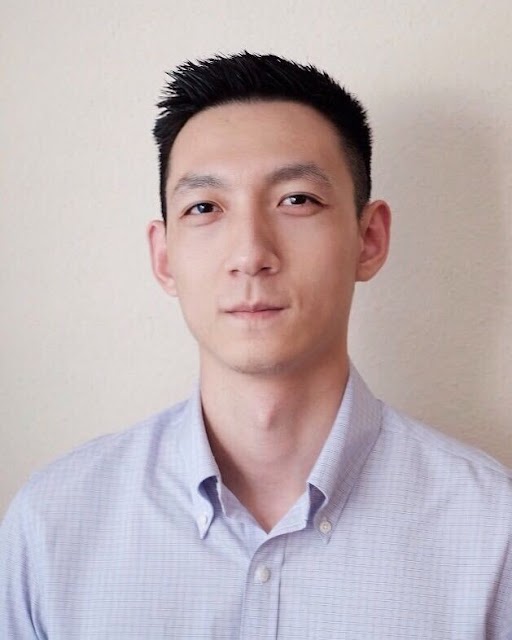}}]{Zhanghao Yu} received a B.E. degree in Integrated Circuit Design and Integrated System from the University of Electronic Science and Technology of China, Chengdu, China, in 2016, and an M.S. degree in Electrical Engineering from the University of Southern California, Los Angeles, CA, in 2018. He is currently pursuing a Ph.D. degree in Electrical and Computer Engineering at Rice University, Houston, TX, advised by Professor Kaiyuan Yang.

His research interests include analog and mixed-signal integrated circuit design for wireless bioelectronics, power management, low-power communication, and security. He is the recipient of the 2021-2022 IEEE Solid-State Circuits Society (SSCS) Predoctoral Achievement Award. As the (co-)first author, he received the Best Paper Award at 2021 IEEE Custom Integrated Circuits Conference (CICC), the Best Paper Award at 28th Annual International Conference on Mobile Computing and Networking (MobiCom), and the Best Student Paper Finalist at 2022 IEEE Radio Frequency Integrated Circuits Symposium (RFIC). 
\end{IEEEbiography}

\begin{IEEEbiography}[{\includegraphics[width=1in,height=1.25in,clip,keepaspectratio]{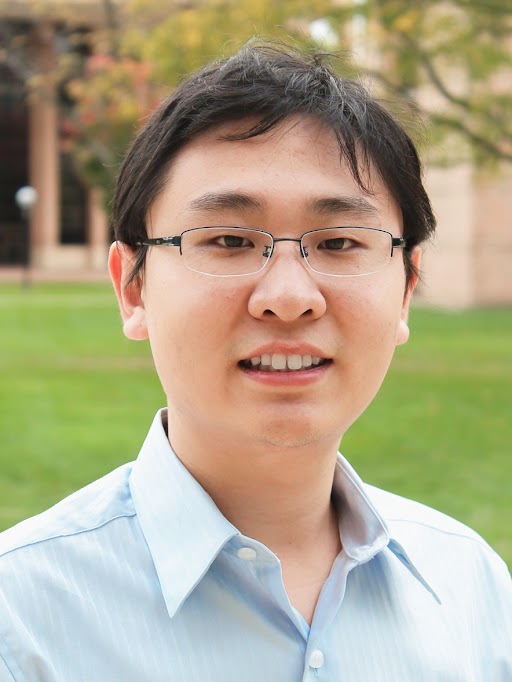}}]{Kaiyuan Yang} received his B.S. in Electronics Engineering from Tsinghua University, Beijing, China, in 2012, and his Ph.D. degree in Electrical Engineering from the University of Michigan, Ann Arbor, MI, in 2017. His research interests include low-power integrated circuit and system design for secure and intelligent microsystems, bioelectronics, hardware security, and mixed-signal computing.

Dr. Yang is currently an Assistant Professor of Electrical and Computer Engineering at Rice University, USA, and leads the Secure and Intelligent Micro-Systems (SIMS) lab. He is a recipient of 2022 National Science Foundation (NSF) CAREER Award and 2016 IEEE Solid-State Circuits Society (SSCS) Predoctoral Achievement Award. He also received best paper awards from premier conferences in various fields, including 2022 ACM Annual International Conference on Mobile Computing and Networking (MobiCom), 2021 IEEE Custom Integrated Circuit Conference (CICC), 2016 IEEE International Symposium on Security and Privacy (Oakland), and 2015 IEEE International Symposium on Circuits and Systems (ISCAS), and multiple best paper award nominations. His research was also recognized as the cover of Nature Biomedical Engineering, the research highlight at Communications of ACM, and Top Picks in Hardware and Embedded Security. He is currently serving as an associate editor of IEEE Transactions on VLSI Systems (TVLSI), the co-chair of IEEE SSCS Houston chapter, and a TPC member of multiple international conferences. 
\end{IEEEbiography}



\vfill


\end{document}